\documentclass[reprint,superscriptaddress,amsmath,amssymb,aps,showkeys,pra,floatfix]{revtex4-2}

\usepackage{graphicx}
\usepackage{dcolumn}
\usepackage{bm}
\usepackage{float}
\usepackage{siunitx}

\begin{document}

\preprint{APS/123-QED}

\title{Experimental observation of self-frequency-shifting Raman quasi-solitons in fiber Fabry-Perot resonators}

\author{Thomas Bunel}
\author{Matteo Conforti}%
\author{Arnaud Mussot}
 \email{arnaud.mussot@univ-lille.fr}
\affiliation{%
University of Lille, CNRS, UMR 8523-PhLAM Physique des Lasers, Atomes et Molécules, F-59000, Lille, France
}%

\date{\today}

\begin{abstract}
We report the generation of self-frequency-shifting Raman quasi-solitons in a pulse-pumped high-Q fiber Fabry-Perot resonator in the weak normal dispersion regime. They are induced by modulation instability mediated by fourth-order dispersion in a regime where the Raman gain overwhelms the cavity losses. The resulting spectrum, spanning over 50~THz, is reminiscent of the supercontinuum generated in single-pass waveguides. For the first time to our knowledge, we clearly identify this process using a dispersive Fourier transform experiment. Additionally, we demonstrate the suppression of modulation instability by tuning the synchronization mismatch between the pump repetition rate and the cavity roundtrip time, enabling the generation of a standard dissipative Kerr soliton in this system. These observations align remarkably well with numerical simulations based on a generalized Lugiato-Lefever equation, 
incorporating the Raman response of the optical fiber.
\end{abstract}

\maketitle


\section{Introduction}
Kerr-type nonlinear resonators have attracted significant attention over the last decade, primarily due to their ability to generate broad and stable optical frequency combs (OFC)~\cite{kippenberg_dissipative_2018,pasquazi_micro-combs_2018,sun_applications_2023}.
Their appeal lies in the fact that they naturally inherit phase stabilization from the pump laser, eliminating the need for complex mode-locking mechanisms and suppressing amplified spontaneous emission noise, as the resonators are purely passive. Significant efforts have been devoted to understanding the dynamics of these systems. Today, the Lugiato-Lefever equation (LLE) is widely used to study Kerr cavity dynamics both theoretically and numerically. In particular, the anomalous dispersion regime has been extensively investigated, from modulation instability (MI)~\cite{coen_continuous-wave_2001,bessin_real-time_2019,Copie:17,negrini_pump-cavity_2023} to optical rogue waves~\cite{Coulibaly_turbulence_2019,Coillet_optical_2014}, as it enables cavity soliton generation, short and intense pulses circulating without deformation in the cavity, resulting in a stable OFC at the cavity output~\cite{kippenberg_dissipative_2018,herr_temporal_2014,brasch_photonic_2016,englebert_high_2023}. Experimentally, various optical systems have been employed to study and generate cavity solitons, including fiber ring cavities~\cite{li_experimental_2020,englebert_temporal_2021,englebert_high_2023}, whispering-gallery-mode resonators~\cite{herr_temporal_2014}, integrated chip microresonators~\cite{brasch_photonic_2016}, or fiber Fabry-Perot (FFP) resonators~\cite{obrzud_temporal_2017,bunel_28_2024}. Depending on the structure and materials of the resonator, additional nonlinear effects such as Brillouin scattering~\cite{lucas_dynamic_2023,nie_dissipative_2022,bunel2025brillouininducedkerrfrequencycomb} or Raman scattering~\cite{Li_Chen_Wu_2025,li_experimental_2023} must be considered. The latter has attracted particular attention due to its significant impact on cavity soliton generation. Although some studies have used the Raman effect to generate ultrashort cavity solitons by pumping the resonator into the normal dispersion regime~\cite{li_ultrashort_2023}, adjusting the repetition rate of the generated OFC~\cite{suh_soliton_2018}, or generating Stokes solitons~\cite{Yang_Yi_Yang_Vahala_2016}, Raman scattering is more commonly known for its detrimental effects on cavity solitons. It limits their bandwidth and duration due to the Raman-induced frequency shift, which causes solitons to drift in the cavity relative to the driving field~\cite{englebert_high_2023,wang_stimulated_2018}. 

First theoretically predicted by C. Milián~\textit{et al.} in 2015~\cite{milian_solitons_2015}, these frequency-locked Raman (FLR) cavity solitons were experimentally observed one year later~\cite{karpov_raman_2016}. However, in their paper, Milián \textit{et al.} also predicted the existence of self-frequency-shifting Raman (SFSR) quasi-solitons in high-quality factor silica microresonators. Unlike FLR solitons, which require a balance between intracavity losses and Raman gain, these Raman-shifting quasi-solitons are generated when intracavity losses are negligible compared to Raman gain. In this case, the LLE can be simplified to the nonlinear Schrödinger equation, which admits SFSR soliton solutions~\cite{milian_solitons_2015}. Consequently, solitons emerging from MI do not have a fixed velocity like FLR solitons but instead undergo continuous acceleration, leading to a persistent shift towards lower frequencies, similar to those observed in single-pass optical fibers, nanophotonic waveguides~\cite{Supercontinuum_2010,dudley_supercontinuum_2006,bres_supercontinuum_2023}, or fiber laser cavities~\cite{Meng_2021}. As with supercontinuum generation in anomalous dispersion waveguides, the dynamics of these solitons are complex and strongly influenced by the dispersive waves they emit~\cite{dudley_supercontinuum_2006}. The key requirements for generating such signals are a high-quality factor resonator (i.e. very low internal losses) with high Raman gain. These conditions significantly constrain the platforms where this phenomenon can be observed. Integrated microresonators exhibit very high finesse, but their narrow Raman gain limits its impact on the soliton spectrum and dynamics. Conversely, fiber-ring cavities provide broad Raman gain, but their picosecond pulse durations and low-finesse factors make the generation of SFSR quasi-solitons impractical. Nonetheless, some experimental studies appear to have observed this type of signal without explicitly identifying it. For instance, in early studies on silica toroidal microcavities~\cite{delhaye_optical_2007,papp_microresonator_2014,li_low-pump-power_2012} and more recently in FFP cavities~\cite{xiao_near-zero-dispersion_2023}. In all these works, the generated spectra closely resemble those predicted in Ref~\cite{milian_solitons_2015}. However, rather than being clearly identified as SFSR quasi-solitons, they were attributed to multiparametric processes~\cite{delhaye_optical_2007,papp_microresonator_2014,li_low-pump-power_2012} enhanced by four-wave mixing or broadband MI~\cite{xiao_near-zero-dispersion_2023}.

In this paper, we specifically investigate SFSR quasi-solitons generation in a fiber Fabry-Perot resonator. FFP resonators are particularly well-suited due to their high-quality factor, which arises from highly reflective mirrors and low-loss optical fiber, as well as the broadband Raman response characteristic of silica fibers~\cite{govind_p_agrawal_non_2013}. In addition to providing a clear experimental observation of SFSR quasi-solitons, we demonstrate that this phenomenon can occur in the normal dispersion regime, where quasi-solitons emerge from modulation instability induced by fourth-order dispersion. By pumping the FFP resonator with a pulse train, we show that either frequency-locked cavity solitons or SFSR quasi-solitons can be generated, depending on the synchronization mismatch between the pump repetition rate and the cavity length. We experimentally investigate the dynamics of the generated SFSR quasi-solitons using the dispersive Fourier transform method.  

\section{\label{sec:setup}Experimental setup}

\begin{figure}
\includegraphics{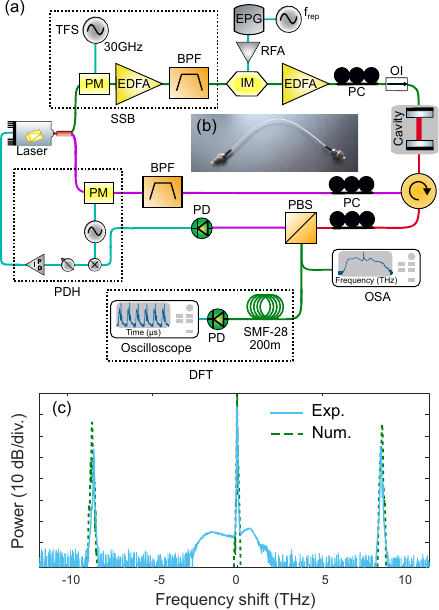}
\caption{\label{fig:setup}Experimental setup and MI-$\beta_4$ spectrum. (a)~Experimental setup with a two-arms stabilization system. Green (upper) line: nonlinear beam; pink (lower) line: control beam. Both beams are perpendicularly polarized to each other. TFS: Tunable Frequency Synthesizer; EPG: Electrical Pulse Generator; RFA: Radio Frequencies Amplifier; IM: Intensity Modulator; PM: Phase Modulator; EDFA: Erbium Doped Fiber Amplifier; PC: Polarization Controller; OI: Optical Isolator; PD: Photodiode; PBS: Polarization Beam Splitter; PDH: Pound-Drever-Hall; SSB: single-side-band generator; ESA: Electrical Spectrum Analyser; OSA: Optical Spectrum Analyser; OSO: Optical Sampling Oscilloscope. (b)~Photograph of the FFP resonator. (c)~Spectrum resolution from MI-$\beta_4$ for $\delta_0=-0.035$~rad, $\Delta T=72$~fs, and $P_{in}=8$~W.}
\end{figure}

The FFP resonator used for this investigation is depicted in Fig.~\ref{fig:setup}(b). It is composed of a 20.72~cm dispersion-shifted fiber with a nonlinear coefficient $\gamma = 2.5$~W$^{-1}$km$^{-1}$. At the pump wavelength of 1550~nm, the fiber operates in the normal dispersion regime, with group velocity dispersion (GVD) $\beta_2=0.145$~ps$^2$km$^{-1}$, third-order dispersion (TOD) $\beta_3=0.115$~ps$^3$km$^{-1}$, and fourth-order dispersion (FOD) $\beta_4=-9.5\times10^{-4}$~ps$^4$km$^{-1}$. The resonator is formed by connecting the optical fiber between two FC/PC connectors, where dielectric mirrors are deposited on each end using a physical vapor deposition technique, achieving a reflectance of $99.86\%$ over a $100$~nm bandwidth around the pump wavelength (1550~nm)~\cite{zideluns_automated_2021}. The free spectral range (FSR) is $496.7$~MHz, and the finesse is $800$, corresponding to a Q-factor of $311.7$~million. This FFP cavity is integrated into the experimental setup depicted in Fig.~\ref{fig:setup}(a), which closely resembles those used in Refs.~\cite{bunel_broadband_2024,bunel_impact_2023}. The setup consists of two optical paths: the pink arm (bottom), which stabilizes the laser against cavity fluctuations, caused by vibrations or temperature variations, using a Pound-Drever-Hall (PDH) locking system~\cite{black_introduction_2001}, and the green arm (top) which pumps the cavity to generate nonlinear effects at a precise detuning value by using an optical pulse train. A homemade single-sideband (SSB) generator is implemented to adjust the detuning as needed, followed by an intensity modulator driven by an electrical pulse generator (EPG) to transform the continuous-wave (CW) pump into a Gaussian pulse train with a duration of $55$~ps. The synchronization between the pump and the cavity is controlled by a frequency synthesizer that determines the pulse repetition rate. Finally, an erbium-doped fiber amplifier (EDFA) boosts the pump peak power to $8$~W. The two beams, one for stabilization (pink) and one for nonlinear generation (green), are orthogonally polarized to minimize energy exchange within the cavity. They are then separated at the cavity output using a polarization beam splitter (PBS). The stabilization signal is sent to the PDH system, while the nonlinear signal is analyzed using an optical spectrum analyzer (OSA), and the dispersive Fourier transform (DFT) measurement technique which will be detailed in Section~\ref{sec:dft}.  

\section{\label{sec:SFSR}Frequency locked solitons and SFSR quasi-solitons}

By setting the cavity detuning to $\delta=-0.035$~rad, and slightly adjusting the pump frequency value to introduce a synchronization mismatch between the pump repetition rate $f_{rep}$ and the cavity roundtrip time, given by $\Delta T = \frac{1}{f_{rep}}-\frac{1}{FSR}=72$~fs, two sidebands are generated at $8.44$~THz from the pump [Fig.~\ref{fig:setup}(c)]. 
This result can be modeled using a generalized LLE adapted for Fabry-Perot resonators, accounting for the additional phase shift induced by counterpropagating waves in such resonators~\cite{firth_analytic_2021,cole_theory_2018,ziani_theory_2024}, as well as the Raman effect~\cite{wang_stimulated_2018}. The generalized LLE reads as follows:
\begin{widetext}
\begin{align}\label{eq:FP-LLE_Raman}
    \nonumber  &2L \frac{\partial \psi}{\partial z} = -\alpha \psi + \theta \sqrt{P_{in}} +i \biggl( -\delta +2L \sum _{n=2}^{n=4} i^n \frac{\beta_n}{n!} \frac{\partial^n}{\partial \tau^n} \biggl) \psi(z,\tau) \\
    & \hspace*{1.5cm} +i\gamma 2L \biggl( (1-f_R)|\psi|^2 + f_R \int_{-\infty}^{\infty} h_R(\tau') |\psi(z,\tau-\tau')|^2 d\tau' + \frac{\chi G}{t_R} \int_{-t_R/2}^{t_R/2} |\psi(z,\tau')|^2 d\tau' \biggl)\psi(z,\tau) 
\end{align}
\end{widetext}
where $\psi$ is the field envelope, $P_{in}=8$~W is the input power, $L=20.72$~cm is the cavity length, $\theta=0.0374$ is the transmissivity of the mirror, $\delta$ is the cavity detuning, $\gamma=2.5$W$^{-1}$km$^{-1}$ is the nonlinear coefficient, $\alpha$ accounts for the total cavity losses, directly linked to the finesse $\mathcal{F}$: $\alpha=\pi/\mathcal{F}=\pi/800$ (valid for $\mathcal{F}\gg1$), $\beta_n~(n \geq 2)$ account for the GVD (for n=2) and higher order dispersion (HOD), $z$ is the longitudinal coordinate, and $\tau$ is the time defined in a reference frame that travels at the group velocity of light in the fiber, and $t' \in [-t_R/2, t_R/2]$ denotes the fast time in one cavity roundtrip, $G=2$, accounts for XPM, and $\chi=0.027$, accounts for the ratio between the pulse duration and the cavity roundtrip time. The parameter $\beta_1=-\frac{\Delta T}{L}$ accounts for the synchronization mismatch between the pump and the cavity, $f_R$ is the fractional contribution of the delayed Raman response, and
\begin{equation}
    h_R(t)=\frac{\tau_1^2+\tau_2^2}{\tau_2^2} \text{exp}(-t/\tau_2) \text{sin}(t/\tau_1)
\end{equation}
is the approximate Raman response function in silica fiber, where $\tau_1=12.2$~fs, $\tau_2=32$~fs, and $f_R$=0.18~\cite{govind_p_agrawal_non_2013}. 
The experimental results [blue line in Fig.~\ref{fig:setup}(c)] show excellent agreement with numerical simulations [green dashed line]. 
We identify this process with modulation instability. Although the cavity operates in the normal dispersion regime, MI arises in the monostable regime due to the negative FOD term, a phenomenon also referred to as MI-$\beta_4$~\cite{bessin_modulation_2017,sayson_octave-spanning_2019}. 
It has been shown that MI frequency can be theoretically calculated by performing a linear stability analysis of Eq.~(\ref{eq:FP-LLE_Raman})~\cite{ziani_theory_2024,bessin_modulation_2017}. The theoretical angular frequency shift of MI-$\beta_4$ can thus be read as follows:
\begin{equation}
\Omega_{MI-FP}=\sqrt{\frac{-6\beta_2 \pm 6 \sqrt{\beta_2^2-\frac{2}{3} \beta_4 \left((2+\chi G) \gamma P -\frac{\delta_0}{2L} \right)}}{\beta_4}}
\label{eq:MIbeta4_FP}
\end{equation}
Theoretical prediction [red arrows in Fig.~\ref{fig:setup}(c)] obtained with Eq.~(\ref{eq:MIbeta4_FP}) predicts MI frequency at $\Omega_{MI-FP}/(2\pi)=8.5167$~THz, in very good agreement with numerics and experimental results. This confirms that synchronization mismatch is necessary to generate the MI process at this specific detuning value. The explanation lies in the convective nature of the resonator. The odd higher-order dispersion and Raman terms introduce nonlocal effects that break the reflection symmetry in the governing equation~(\ref{eq:FP-LLE_Raman}), leading to convective instabilities~\cite{mussot_optical_2008,leo_nonlinear_2013}. As a result, the MI-induced patterns drift relative to the driving field. In the case of CW pumping, this drift is not problematic, as the cavity is continuously "filled" allowing MI to develop while still being sustained by the pump. However, in a pulse pumping scheme, the generated patterns drift until they reach the edge of the driving pulse, at which point they vanish due to losses if the pump is perfectly synchronized with the cavity. By introducing a synchronization mismatch of $\Delta T=72$~fs, corresponding to a pump group velocity shift of $-\Delta T/(2L)=-175$~ps$\cdot$km$^{-1}$, MI can be maintained within the driving pulse.

\begin{figure}
\includegraphics{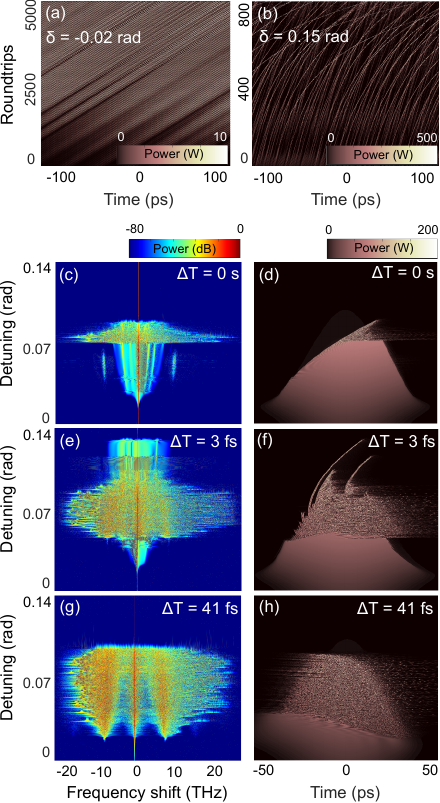}
\caption{\label{fig:simus}Numerical simulations of FFP cavity nonlinear dynamics. (a) and (b)~Simulation of the propagation of a CW pumped FFP resonator showing MI-$\beta_4$ process (a) and SFSR quas-solitons generation. (c)-(h)~Nonlinear resonance scan as a function of pump synchronisation mismatch, obtained with numerics. (c), (e) and (g)~evolution of time domain signal with detuning. (d), (f) and (h)~evolution of frequency domain signal with detuning. Parameters: $L=20.72$~cm, $\beta_2=0.145$~ps$^2$km$^{-1}$, $\beta_3=0.115$~ps$^3$km$^{-1}$, $\beta_4=-9.5\times10^{-4}$~ps$^4$km$^{-1}$, $\gamma=2.5$~W$^{-1}$km$^{-1}$, $FSR=496.7$~MHz, $\mathcal{F}=800$, $P_{in}=5$~W.}
\end{figure}

Further numerical simulations presented in Fig.~\ref{fig:simus} provide deeper insight into the cavity dynamics. Fig.~\ref{fig:simus}(a) illustrates a propagation simulation in the time domain as a function of the number of cavity roundtrips,
at a detuning of $\delta=-0.02$~rad when the cavity is CW pumped. The convective instability is evident from the drift of the MI-induced patterns relative to the driving field, with a group velocity offset of $178$~ps$\cdot$km$^{-1}$. In a very similar manner to the numerical simulations reported in Ref.~\cite{milian_solitons_2015}, increasing the cavity detuning leads to the generation of SFSR quasi-solitons. Specifically, for the same parameters but with an increased detuning of $\delta=0.15$~rad, we observe that MI generates localized pulses that experience continuous acceleration relative to the pump field [Fig.~\ref{fig:simus}(b)]. 

We now consider the pulsed pump case. Fig.~\ref{fig:simus}(c)-(f) provides a brief overview of the different dynamics depending on the synchronization mismatch $\Delta T$. 
The group velocity offset of a localized structure depends on its power and frequency shift~\cite{mussot_optical_2008,milian_soliton_2014}. It can thus be expected that cavity solitons do not travel at the same speed as MI-$\beta_4$~\cite{li_experimental_2020}. To explore this phenomenon, we use Eq.~(\ref{eq:FP-LLE_Raman}) to simulate cavity scans with a scanning speed of $3\times 10^{-6}$~rad/roundtrip,
and varying synchronization mismatch $\Delta T$. In Fig.~\ref{fig:simus}(c)-(h), the evolution of the time-domain and frequency-domain signals is shown for three different $\Delta T$ values. It is evident that the process strongly depends on the synchronization mismatch. When the pump is perfectly synchronized with the cavity ($\Delta T=0$), the nonlinear process mainly results in spectral broadening near the pump, driven by switching wave formation due to the positive GVD value~\cite{macnaughtan_temporal_2023,bunel_broadband_2024}. Weak MI bands are generated around $\delta_0=0.03$~rad [Fig.~\ref{fig:simus}(c)], before vanishing due to their drift [Fig.~\ref{fig:simus}(d)]. When the desynchronization is increased to $\Delta T=3$~fs, a powerful incoherent nonlinear regime is observed between $0.04$ and $0.08$~rad [Fig.~\ref{fig:simus}(e) and (h)], followed by the emergence of solitons that persist until the detuning reaches $0.13$~rad. Therefore, to generate frequency-locked cavity solitons, the pump must have an adequate synchronization mismatch to travel with the soliton circulating in the cavity. This conclusion closely aligns with previous studies on cavity soliton generation in weak dispersion Kerr resonators~\cite{anderson_zero_2022,xiao_near-zero-dispersion_2023}. Finally, when a strong pump-to-cavity synchronization mismatch is applied ($\Delta T=41$~fs), strong MI sidebands appear at $\delta=0.008$~rad [Fig.~\ref{fig:simus}(g)], as the pump now follows the MI drift. Note that MI-$\beta_4$ actually emerges for lower values of the detuning, but it is not visible in this instance as it requires more time to reach a steady state at low detuning. This allows self-frequency-shifting Raman quasi-solitons to be generated when detuning is increased. From $\delta=0.02$~rad to $0.1$~rad, we observe a high spectral broadening with significant intensity. Localized frequency shifts towards lower frequencies (around $-15$~THz compared to the pump) serve as signatures of SFSR quasi-solitons.
To resume, in a pulse pumping scheme, the pump repetition rate must be desynchronized from the cavity FSR to allow MI to persist. 
In this manner, it is possible to generate SFSR quasi-solitons which emerge from MI. Additionally, by adjusting the pump synchronization mismatch, frequency-locked solitons can also be generated. This reveals the ability to generate both types of localized patterns, one leading to a stable Kerr frequency comb (frequency-locked solitons) and the other to a broadband supercontinuum (SFSR quasi-solitons), by simply changing this linear parameter.

\begin{figure*}
\includegraphics{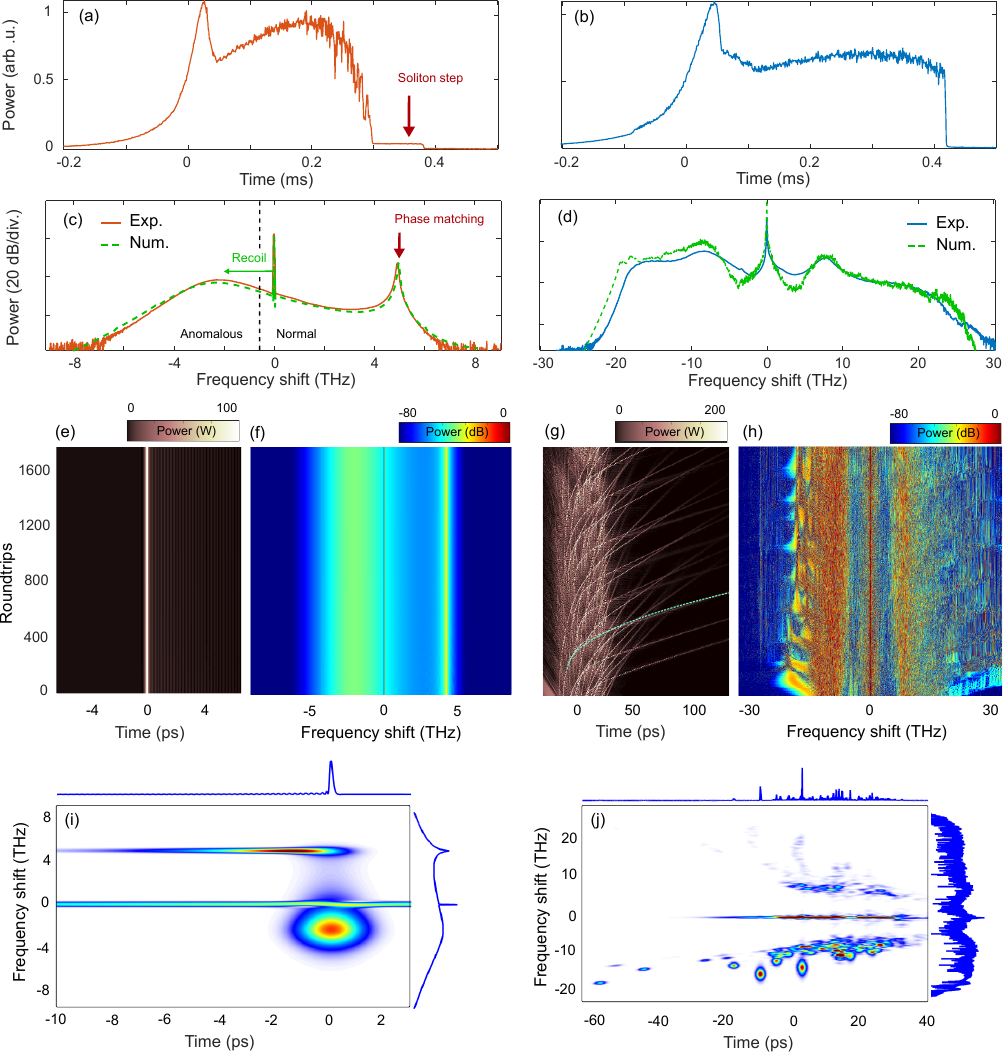}
\caption{\label{fig:exp}Cavity soliton-induced OFC and SFSR quasi-solitons-induced supercontinuum. (a), (c), (e), (f) and (i) are numerical and experimental results for frequency-locked cavity solitons; and (b), (d), (g), (h) and (j) concern SFSR quasi-solitons (a) and (b)~Experimental nonlinear transfer function recording with a pump frequency sweep at 0.8~GHz/s. (c) and (d)~Experimental and simulated spectra. (e) and (g)~Time domain signal evolution. (f) and (h)~Frequency domain signal evolution. (i) and (j)~Spectrogram of an output pulse. (e) to (j) are obtained through numerics.}
\end{figure*}

Experimentally, it is therefore possible to transition from frequency-locked cavity soliton generation to SFSR quasi-soliton generation simply by adjusting the synchronization mismatch $\Delta T$, which corresponds to modifying the frequency synthesizer value that delivers $f_{rep}$ [see Fig.~\ref{fig:setup}(a)].
Setting $\Delta T = 3$~fs, which corresponds to set the pump frequency to $f_{rep}=FSR-740$~Hz, and performing a resonance scan from blue to red detuning, the nonlinear transfer function, shown in Fig.~\ref{fig:exp}(a), exhibits a characteristic step corresponding to cavity soliton formation. Note that experimentally, the pump repetition rate is gradually decreased while continuously scanning the cavity resonances until the soliton steps become visible. By tuning the SSB frequency value [see Fig.~\ref{fig:setup}(a)] to match the detuning associated with this soliton step, single cavity solitons can be generated, as shown in Fig.~\ref{fig:exp}(c). The experimental results [orange line] are in excellent agreement with numerical simulations obtained by solving Eq.~(\ref{eq:FP-LLE_Raman}) for a cavity detuning of $\delta_0=0.11$~rad [green dashed line]. As expected, a strong dispersive wave (DW) is generated due to TOD, causing the soliton to drift in the cavity reference frame. This drift in the temporal domain is accompanied by a spectral recoil [green arrow in Fig.~\ref{fig:exp}(c)], which allows the soliton to remain stable and in the anomalous dispersion regime despite being pumped in the normal dispersion region~\cite{li_experimental_2020,zhang_quintic_2023}. The position of the DW can be determined using the phase-matching condition~\cite{milian_soliton_2014,conforti_dispersive_2013,jang_observation_2014}, given by:
\begin{equation}
\label{eq:phaseMatching}
\frac{2L\beta_3}{6}\omega^3+\frac{2L\beta_2}{2}\omega^2-D\omega-\delta=0
\end{equation}
for which the nonlinear contribution has been neglected~\cite{bunel_28_2024}. Here, $\omega$ is the normalized angular frequency shift of the driving field, and $D$ represents the group-delay accumulated by the temporal cavity soliton with respect to the cavity frame over one roundtrip (in units of time). The spectral recoil is measured to be $\Omega_{CS}/(2 \pi)=2.32$~THz, corresponding to a soliton group delay of $D = 2L\beta_2 \Omega_{CS} + 2L\beta_3 \Omega_{CS}^2/2 =1.73$~fs in good agreement with the pump synchronization mismatch $\Delta T=3$~fs.
Thus, the phase-matching condition [red arrow in Fig.~\ref{fig:exp}(b)] perfectly matches the peak position corresponding to the DW emission (4.91~THz predicted by Eq.~(\ref{eq:phaseMatching}) and 4.82~THz observed). As highlighted in the spectrogram in Fig.~\ref{fig:exp}(i), the DW results in an oscillatory tail traveling with the soliton~\cite{jang_observation_2014}. This process is stable and do not change with time once generated. Time domain pattern and spectrum repeat roundtrip after roundtrip as reported in Fig.~\ref{fig:exp}(e) and (f), corresponding to an OFC generation at the cavity output. Note that the drift of the cavity soliton is not visible in Fig.~\ref{fig:exp}(e) as it is plotted in the reference frame of the pump which is desynchronized compare to the cavity as reported above.

The scenario changes drastically when the synchronization mismatch is set to $\Delta T=72$~fs ($f_{rep}=FSR-17.763$~kHz), enabling the generation of SFSR quasi-solitons. In this case, no soliton step appears in the nonlinear transfer function [Fig.~\ref{fig:exp}(b)]. By locking the system at a cavity detuning of $\delta=0.08$~rad, we observe significant spectral broadening [blue line in Fig.~\ref{fig:exp}(d)]. While the MI-$\beta_4$ lobes remain visible around $\pm 8$~THz from the pump, the generated spectrum extends much further, spanning over 50~THz. This confirms the generation of SFSR quasi-solitons, akin to those predicted in Ref.~\cite{milian_solitons_2015}. Numerics obtained with Eq.~(\ref{eq:FP-LLE_Raman}) reveal the formation of ultrashort pulses, on the order of a few tens of femtoseconds, within the 55~ps pump pulse [Figure~\ref{fig:exp}(g)]. These pulses undergo constant acceleration before exiting the pump pulse and eventually disappearing. As an example, the light blue dashed line outlines a quasi-soliton acceleration of $12.1$~ns$\cdot$m$^{-2}$, with a maximum velocity of $0.9$~ps$\cdot$m$^{-1}$. These SFSR quasi-solitons originate from MI and subsequently shift toward lower frequencies due to the Raman effect during their acceleration, while emitting dispersive waves at phase-matched frequencies. This behavior closely resembles soliton self-frequency shift and soliton fission in single-pass nonlinear waveguides~\cite{dudley_supercontinuum_2006,Supercontinuum_2010}. Further evidence of this phenomenon is provided in Fig.~\ref{fig:exp}(j), which presents the spectrogram of a simulated output pulse. The SFSR quasi-solitons [circular patterns] shift toward lower frequencies (between $-15$ and $-20$~THz) due to the Raman effect, while emitting dispersive waves at higher frequencies where phase matching occurs [blue spots between $15$ and $30$~THz]. The spectral variations over multiple cavity roundtrips [Fig.~\ref{fig:exp}(h)] illustrate this dynamic process, characterized by significant emissions in the Stokes side when the SFSR quasi-soliton reaches its maximum velocity [red spots around $-15$~THz and $-20$~THz], accompanied by high-frequency radiation [blue traces between $15$ and $30$~THz]. However, this phenomenon differs from single-pass pulse propagation in optical fibers because it is continuously repeated due to the pump replenishing the cavity. Unlike Kerr frequency combs, the signal evolves with each cavity roundtrip, rendering the output signal incoherent. To accurately simulate the measurement taken with the OSA [blue line in Fig.~\ref{fig:exp}(d)], we average all output signals over a large number of roundtrips—1000 for the numerical result shown in green in Fig.~\ref{fig:exp}(d). The numerical simulations (green dashed lines) exhibit excellent agreement with experimental results. 

\begin{figure*}
\includegraphics{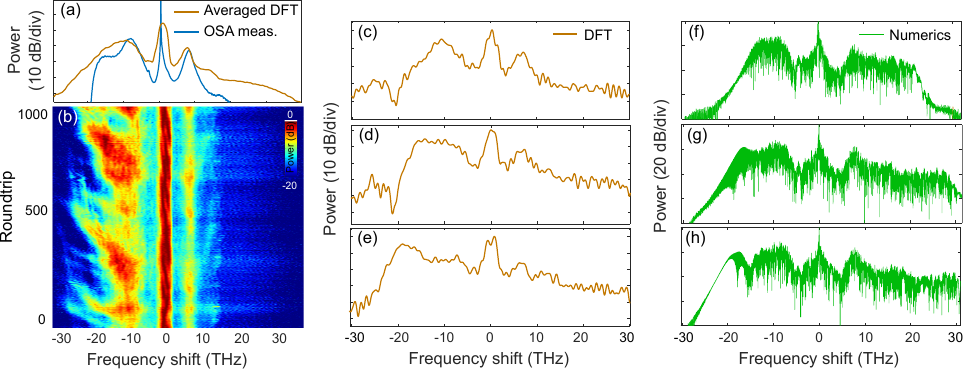}
\caption{\label{fig:dft}DFT experiment. (a)~Comparison between OSA measurement and DFT measurement. (b)~Frequency domain signal evolution, obtained through DFT. (c), (d) and (e)~Single shot measurement with DFT during a quasi-soliton Raman self-frequency shift. (f), (g) and (h)~Corresponding numerics.}
\end{figure*}

\section{\label{sec:dft}Observation of SFSR quasi-solitons with DFT experiment}

To clearly identify the generation of SFSR quasi-soliton in a Kerr resonator, we conduct a dispersive Fourier Transform (DFT) experiment~\cite{goda_dispersive_2013,godin_recent_2022} to record the spectrum roundtrip by roundtrip and observe its evolution. DFT, also known as real-time Fourier transformation, utilizes the analogy between paraxial diffraction and temporal dispersion~\cite{kolner_space-time_1994} to map the temporal frequency spectrum of a pulse to a temporal waveform, whose intensity envelope mimics the spectrum. The method involves using a purely dispersive element with a large group velocity dispersion and a photodetector. When an optical pulse train enters the dispersive element, the spectrum of each pulse is mapped to a temporal waveform due to the large GVD in the dispersive element. The temporally dispersed pulse train is then detected by the photodetector and subjected to digital signal processing for pulse-by-pulse spectral analysis. This simple method has been employed in practical applications such as spectroscopy~\cite{chou_real-time_2008}. More relevant to our system, DFT has been applied for dynamic studies on supercontinuum generation and OFC formation as well~\cite{solli_active_2008,wetzel_real-time_2012,cutrona_nonlocal_2023,lapre_dispersive_2020,Copie_mod_2017}.

For our experiment, we use 200~m of SMF-28 as a dispersive element [Fig.~\ref{fig:setup}(a)]. The temporal trace mimics the spectrum, and the corresponding frequency can be retrieved using the relation:
\begin{equation}
    t(\omega) = \sum_{n=2} \frac{\beta_n z}{(n-1)!} (\omega-\omega_0)^{n-1}
\end{equation}
where $t$ is the time, $z$ is the propagation length, $\omega$ is the angular frequency, and $\omega_0$ accounts for the pump frequency. Since the second-order dispersion of the SMF-28 fiber dominates over higher-order dispersions, we simplify this expression as $\omega=\omega_0 + \frac{t(\omega)}{\beta_2 z}$. Using a photodiode and oscilloscope, we plot a map representing the evolution of the spectrum as a function of cavity roundtrips, as shown in Fig.~\ref{fig:dft}(b). To verify the validity of the measurement, we average all the spectra recorded by DFT and compare the result to the OSA measurement in Fig.~\ref{fig:dft}(a). It shows good agreement, particularly in the position of the MI bands. Similar to the numerical results [Fig.~\ref{fig:exp}(j)], we observe localized increases around $-20$~THz, accompanied by slight emissions between 10 and 30~THz, indicating the soliton fission phenomenon with Raman shifts and DW emissions.
In Fig.~\ref{fig:dft}(c), (d), and (e), we show examples of the recorded traces during a soliton Raman self-frequency shift [from (c) to (e)], plotted in orange. For comparison, Fig.~\ref{fig:dft}(f), (g), and (h), in green, represent numerical traces,
extracted from the simulation made for Fig.~\ref{fig:exp}(h), corresponding to the same phenomenon. In both cases, the Raman shift progressively moves from the MI bands toward lower frequencies, accompanied by an increase in higher frequencies due to DW emissions. These single-shot numerical and experimental results are in good agreement, providing further evidence of the existence of SFSR quasi-solitons in the FFP resonator. The observed dynamics are very similar to those reported in single-pass configuration nonlinear fibers, showing strong resemblance to the results reported in Ref.~\cite{dudley_supercontinuum_2006}.

\section{Discussion and conclusion}
In this study, we have investigated the nonlinear dynamics of a weak normal dispersion high-Q fiber Fabry-Perot resonator, where strong Raman effects occur. By using a pulsed pump with a tunable repetition rate, we were able to selectively generate different localized structures depending on the synchronization mismatch between the pump repetition rate and the cavity FSR. 
Two types of localized structures have been observed: frequency-locked cavity solitons and self-frequency shifting Raman quasi-solitons. While cavity solitons in the normal dispersion regime have been previously studied in fiber ring cavities and microresonators~\cite{li_experimental_2020,zhang_quintic_2023}, we have demonstrated the generation of such solitons in an FFP cavity. 

The main result of this study is the experimental observation of SFSR quasi-solitons, which were previously only theoretically predicted~\cite{milian_solitons_2015}. The FFP resonator is an ideal device for realizing this regime due to its high-Q factor and the predominance of Raman gain over losses, facilitated by the optical fiber composition. 
Using real-time measurements via the dispersive Fourier transform technique, we successfully demonstrated soliton fission. The results are in good agreement with numerical simulations, obtained through a generalized Lugiato-Lefever equation, which includes the Raman response.
Furthermore, we have shown that this phenomenon occurs in the normal dispersion regime, provided that the fourth-order dispersion is negative. This condition enables the onset of modulation instability, which is essential for SFSR quasi-soliton generation. Although the coherence of this regime is lower than that of cavity soliton frequency combs, the generated spectra span beyond 50~THz.  

This research advances the understanding of soliton dynamics in high-Q Kerr resonators and paves the way for innovative platforms in OFC and supercontinuum generation.  

\begin{acknowledgments}
This work was supported by the Agence Nationale de la Recherche (Programme Investissements d’Avenir, FARCO and VISOPEC projects; European Regional Development Fund (Photonics for Society P4S) and the CNRS (IRP LAFONI).
\end{acknowledgments}

\section*{Data Availability Statement}

The data that support the findings of this study are available from the corresponding author upon reasonable request.

\nocite{*}

\bibliography{references}

\begin{thebibliography}{62}%
\makeatletter
\providecommand \@ifxundefined [1]{%
 \@ifx{#1\undefined}
}%
\providecommand \@ifnum [1]{%
 \ifnum #1\expandafter \@firstoftwo
 \else \expandafter \@secondoftwo
 \fi
}%
\providecommand \@ifx [1]{%
 \ifx #1\expandafter \@firstoftwo
 \else \expandafter \@secondoftwo
 \fi
}%
\providecommand \natexlab [1]{#1}%
\providecommand \enquote  [1]{``#1''}%
\providecommand \bibnamefont  [1]{#1}%
\providecommand \bibfnamefont [1]{#1}%
\providecommand \citenamefont [1]{#1}%
\providecommand \href@noop [0]{\@secondoftwo}%
\providecommand \href [0]{\begingroup \@sanitize@url \@href}%
\providecommand \@href[1]{\@@startlink{#1}\@@href}%
\providecommand \@@href[1]{\endgroup#1\@@endlink}%
\providecommand \@sanitize@url [0]{\catcode `\\12\catcode `\$12\catcode `\&12\catcode `\#12\catcode `\^12\catcode `\_12\catcode `\%12\relax}%
\providecommand \@@startlink[1]{}%
\providecommand \@@endlink[0]{}%
\providecommand \url  [0]{\begingroup\@sanitize@url \@url }%
\providecommand \@url [1]{\endgroup\@href {#1}{\urlprefix }}%
\providecommand \urlprefix  [0]{URL }%
\providecommand \Eprint [0]{\href }%
\providecommand \doibase [0]{https://doi.org/}%
\providecommand \selectlanguage [0]{\@gobble}%
\providecommand \bibinfo  [0]{\@secondoftwo}%
\providecommand \bibfield  [0]{\@secondoftwo}%
\providecommand \translation [1]{[#1]}%
\providecommand \BibitemOpen [0]{}%
\providecommand \bibitemStop [0]{}%
\providecommand \bibitemNoStop [0]{.\EOS\space}%
\providecommand \EOS [0]{\spacefactor3000\relax}%
\providecommand \BibitemShut  [1]{\csname bibitem#1\endcsname}%
\let\auto@bib@innerbib\@empty
\bibitem [{\citenamefont {Kippenberg}\ \emph {et~al.}(2018)\citenamefont {Kippenberg}, \citenamefont {Gaeta}, \citenamefont {Lipson},\ and\ \citenamefont {Gorodetsky}}]{kippenberg_dissipative_2018}%
  \BibitemOpen
  \bibfield  {author} {\bibinfo {author} {\bibfnamefont {T.~J.}\ \bibnamefont {Kippenberg}}, \bibinfo {author} {\bibfnamefont {A.~L.}\ \bibnamefont {Gaeta}}, \bibinfo {author} {\bibfnamefont {M.}~\bibnamefont {Lipson}},\ and\ \bibinfo {author} {\bibfnamefont {M.~L.}\ \bibnamefont {Gorodetsky}},\ }\bibfield  {title} {\bibinfo {title} {Dissipative {Kerr} solitons in optical microresonators},\ }\href {https://doi.org/10.1126/science.aan8083} {\bibfield  {journal} {\bibinfo  {journal} {Science}\ }\textbf {\bibinfo {volume} {361}},\ \bibinfo {pages} {eaan8083} (\bibinfo {year} {2018})}\BibitemShut {NoStop}%
\bibitem [{\citenamefont {Pasquazi}\ \emph {et~al.}(2018)\citenamefont {Pasquazi}, \citenamefont {Peccianti}, \citenamefont {Razzari}, \citenamefont {Moss}, \citenamefont {Coen}, \citenamefont {Erkintalo}, \citenamefont {Chembo}, \citenamefont {Hansson}, \citenamefont {Wabnitz}, \citenamefont {Del’Haye}, \citenamefont {Xue}, \citenamefont {Weiner},\ and\ \citenamefont {Morandotti}}]{pasquazi_micro-combs_2018}%
  \BibitemOpen
  \bibfield  {author} {\bibinfo {author} {\bibfnamefont {A.}~\bibnamefont {Pasquazi}}, \bibinfo {author} {\bibfnamefont {M.}~\bibnamefont {Peccianti}}, \bibinfo {author} {\bibfnamefont {L.}~\bibnamefont {Razzari}}, \bibinfo {author} {\bibfnamefont {D.~J.}\ \bibnamefont {Moss}}, \bibinfo {author} {\bibfnamefont {S.}~\bibnamefont {Coen}}, \bibinfo {author} {\bibfnamefont {M.}~\bibnamefont {Erkintalo}}, \bibinfo {author} {\bibfnamefont {Y.~K.}\ \bibnamefont {Chembo}}, \bibinfo {author} {\bibfnamefont {T.}~\bibnamefont {Hansson}}, \bibinfo {author} {\bibfnamefont {S.}~\bibnamefont {Wabnitz}}, \bibinfo {author} {\bibfnamefont {P.}~\bibnamefont {Del’Haye}}, \bibinfo {author} {\bibfnamefont {X.}~\bibnamefont {Xue}}, \bibinfo {author} {\bibfnamefont {A.~M.}\ \bibnamefont {Weiner}},\ and\ \bibinfo {author} {\bibfnamefont {R.}~\bibnamefont {Morandotti}},\ }\bibfield  {title} {\bibinfo {title} {Micro-combs: {A} novel generation of optical sources},\ }\href {https://doi.org/10.1016/j.physrep.2017.08.004} {\bibfield
  {journal} {\bibinfo  {journal} {Physics Reports}\ }\textbf {\bibinfo {volume} {729}},\ \bibinfo {pages} {1} (\bibinfo {year} {2018})}\BibitemShut {NoStop}%
\bibitem [{\citenamefont {Sun}\ \emph {et~al.}(2023)\citenamefont {Sun}, \citenamefont {Wu}, \citenamefont {Tan}, \citenamefont {Xu}, \citenamefont {Li}, \citenamefont {Morandotti}, \citenamefont {Mitchell},\ and\ \citenamefont {Moss}}]{sun_applications_2023}%
  \BibitemOpen
  \bibfield  {author} {\bibinfo {author} {\bibfnamefont {Y.}~\bibnamefont {Sun}}, \bibinfo {author} {\bibfnamefont {J.}~\bibnamefont {Wu}}, \bibinfo {author} {\bibfnamefont {M.}~\bibnamefont {Tan}}, \bibinfo {author} {\bibfnamefont {X.}~\bibnamefont {Xu}}, \bibinfo {author} {\bibfnamefont {Y.}~\bibnamefont {Li}}, \bibinfo {author} {\bibfnamefont {R.}~\bibnamefont {Morandotti}}, \bibinfo {author} {\bibfnamefont {A.}~\bibnamefont {Mitchell}},\ and\ \bibinfo {author} {\bibfnamefont {D.~J.}\ \bibnamefont {Moss}},\ }\bibfield  {title} {\bibinfo {title} {Applications of optical microcombs},\ }\href {https://doi.org/10.1364/AOP.470264} {\bibfield  {journal} {\bibinfo  {journal} {Advances in Optics and Photonics}\ }\textbf {\bibinfo {volume} {15}},\ \bibinfo {pages} {86} (\bibinfo {year} {2023})}\BibitemShut {NoStop}%
\bibitem [{\citenamefont {Coen}\ and\ \citenamefont {Haelterman}(2001)}]{coen_continuous-wave_2001}%
  \BibitemOpen
  \bibfield  {author} {\bibinfo {author} {\bibfnamefont {S.}~\bibnamefont {Coen}}\ and\ \bibinfo {author} {\bibfnamefont {M.}~\bibnamefont {Haelterman}},\ }\bibfield  {title} {\bibinfo {title} {Continuous-wave ultrahigh-repetition-rate pulse-train generation through modulational instability in a passive fiber cavity},\ }\href {https://doi.org/10.1364/OL.26.000039} {\bibfield  {journal} {\bibinfo  {journal} {Optics Letters}\ }\textbf {\bibinfo {volume} {26}},\ \bibinfo {pages} {39} (\bibinfo {year} {2001})}\BibitemShut {NoStop}%
\bibitem [{\citenamefont {Bessin}\ \emph {et~al.}(2019)\citenamefont {Bessin}, \citenamefont {Copie}, \citenamefont {Conforti}, \citenamefont {Kudlinski}, \citenamefont {Mussot},\ and\ \citenamefont {Trillo}}]{bessin_real-time_2019}%
  \BibitemOpen
  \bibfield  {author} {\bibinfo {author} {\bibfnamefont {F.}~\bibnamefont {Bessin}}, \bibinfo {author} {\bibfnamefont {F.}~\bibnamefont {Copie}}, \bibinfo {author} {\bibfnamefont {M.}~\bibnamefont {Conforti}}, \bibinfo {author} {\bibfnamefont {A.}~\bibnamefont {Kudlinski}}, \bibinfo {author} {\bibfnamefont {A.}~\bibnamefont {Mussot}},\ and\ \bibinfo {author} {\bibfnamefont {S.}~\bibnamefont {Trillo}},\ }\bibfield  {title} {\bibinfo {title} {Real-{Time} {Characterization} of {Period}-{Doubling} {Dynamics} in {Uniform} and {Dispersion} {Oscillating} {Fiber} {Ring} {Cavities}},\ }\href {https://doi.org/10.1103/PhysRevX.9.041030} {\bibfield  {journal} {\bibinfo  {journal} {Physical Review X}\ }\textbf {\bibinfo {volume} {9}},\ \bibinfo {pages} {041030} (\bibinfo {year} {2019})}\BibitemShut {NoStop}%
\bibitem [{\citenamefont {Copie}\ \emph {et~al.}(2017{\natexlab{a}})\citenamefont {Copie}, \citenamefont {Conforti}, \citenamefont {Kudlinski}, \citenamefont {Trillo},\ and\ \citenamefont {Mussot}}]{Copie:17}%
  \BibitemOpen
  \bibfield  {author} {\bibinfo {author} {\bibfnamefont {F.}~\bibnamefont {Copie}}, \bibinfo {author} {\bibfnamefont {M.}~\bibnamefont {Conforti}}, \bibinfo {author} {\bibfnamefont {A.}~\bibnamefont {Kudlinski}}, \bibinfo {author} {\bibfnamefont {S.}~\bibnamefont {Trillo}},\ and\ \bibinfo {author} {\bibfnamefont {A.}~\bibnamefont {Mussot}},\ }\bibfield  {title} {\bibinfo {title} {Dynamics of turing and faraday instabilities in a longitudinally modulated fiber-ring cavity},\ }\href {https://doi.org/10.1364/OL.42.000435} {\bibfield  {journal} {\bibinfo  {journal} {Opt. Lett.}\ }\textbf {\bibinfo {volume} {42}},\ \bibinfo {pages} {435} (\bibinfo {year} {2017}{\natexlab{a}})}\BibitemShut {NoStop}%
\bibitem [{\citenamefont {Negrini}\ \emph {et~al.}(2023)\citenamefont {Negrini}, \citenamefont {Coulibaly}, \citenamefont {Copie}, \citenamefont {Taki},\ and\ \citenamefont {Mussot}}]{negrini_pump-cavity_2023}%
  \BibitemOpen
  \bibfield  {author} {\bibinfo {author} {\bibfnamefont {S.}~\bibnamefont {Negrini}}, \bibinfo {author} {\bibfnamefont {S.}~\bibnamefont {Coulibaly}}, \bibinfo {author} {\bibfnamefont {F.}~\bibnamefont {Copie}}, \bibinfo {author} {\bibfnamefont {M.}~\bibnamefont {Taki}},\ and\ \bibinfo {author} {\bibfnamefont {A.}~\bibnamefont {Mussot}},\ }\bibfield  {title} {\bibinfo {title} {Pump-cavity synchronization mismatch in modulation instability induced optical frequency combs},\ }\href {https://doi.org/10.1103/PhysRevResearch.5.023133} {\bibfield  {journal} {\bibinfo  {journal} {Physical Review Research}\ }\textbf {\bibinfo {volume} {5}},\ \bibinfo {pages} {023133} (\bibinfo {year} {2023})}\BibitemShut {NoStop}%
\bibitem [{\citenamefont {Coulibaly}\ \emph {et~al.}(2019)\citenamefont {Coulibaly}, \citenamefont {Taki}, \citenamefont {Bendahmane}, \citenamefont {Millot}, \citenamefont {Kibler},\ and\ \citenamefont {Clerc}}]{Coulibaly_turbulence_2019}%
  \BibitemOpen
  \bibfield  {author} {\bibinfo {author} {\bibfnamefont {S.}~\bibnamefont {Coulibaly}}, \bibinfo {author} {\bibfnamefont {M.}~\bibnamefont {Taki}}, \bibinfo {author} {\bibfnamefont {A.}~\bibnamefont {Bendahmane}}, \bibinfo {author} {\bibfnamefont {G.}~\bibnamefont {Millot}}, \bibinfo {author} {\bibfnamefont {B.}~\bibnamefont {Kibler}},\ and\ \bibinfo {author} {\bibfnamefont {M.~G.}\ \bibnamefont {Clerc}},\ }\bibfield  {title} {\bibinfo {title} {Turbulence-induced rogue waves in kerr resonators},\ }\href {https://doi.org/10.1103/PhysRevX.9.011054} {\bibfield  {journal} {\bibinfo  {journal} {Phys. Rev. X}\ }\textbf {\bibinfo {volume} {9}},\ \bibinfo {pages} {011054} (\bibinfo {year} {2019})}\BibitemShut {NoStop}%
\bibitem [{\citenamefont {Coillet}\ \emph {et~al.}(2014)\citenamefont {Coillet}, \citenamefont {Dudley}, \citenamefont {Genty}, \citenamefont {Larger},\ and\ \citenamefont {Chembo}}]{Coillet_optical_2014}%
  \BibitemOpen
  \bibfield  {author} {\bibinfo {author} {\bibfnamefont {A.}~\bibnamefont {Coillet}}, \bibinfo {author} {\bibfnamefont {J.}~\bibnamefont {Dudley}}, \bibinfo {author} {\bibfnamefont {G.}~\bibnamefont {Genty}}, \bibinfo {author} {\bibfnamefont {L.}~\bibnamefont {Larger}},\ and\ \bibinfo {author} {\bibfnamefont {Y.~K.}\ \bibnamefont {Chembo}},\ }\bibfield  {title} {\bibinfo {title} {Optical rogue waves in whispering-gallery-mode resonators},\ }\href {https://doi.org/10.1103/PhysRevA.89.013835} {\bibfield  {journal} {\bibinfo  {journal} {Phys. Rev. A}\ }\textbf {\bibinfo {volume} {89}},\ \bibinfo {pages} {013835} (\bibinfo {year} {2014})}\BibitemShut {NoStop}%
\bibitem [{\citenamefont {Herr}\ \emph {et~al.}(2014)\citenamefont {Herr}, \citenamefont {Brasch}, \citenamefont {Jost}, \citenamefont {Wang}, \citenamefont {Kondratiev}, \citenamefont {Gorodetsky},\ and\ \citenamefont {Kippenberg}}]{herr_temporal_2014}%
  \BibitemOpen
  \bibfield  {author} {\bibinfo {author} {\bibfnamefont {T.}~\bibnamefont {Herr}}, \bibinfo {author} {\bibfnamefont {V.}~\bibnamefont {Brasch}}, \bibinfo {author} {\bibfnamefont {J.~D.}\ \bibnamefont {Jost}}, \bibinfo {author} {\bibfnamefont {C.~Y.}\ \bibnamefont {Wang}}, \bibinfo {author} {\bibfnamefont {N.~M.}\ \bibnamefont {Kondratiev}}, \bibinfo {author} {\bibfnamefont {M.~L.}\ \bibnamefont {Gorodetsky}},\ and\ \bibinfo {author} {\bibfnamefont {T.~J.}\ \bibnamefont {Kippenberg}},\ }\bibfield  {title} {\bibinfo {title} {Temporal solitons in optical microresonators},\ }\href {https://doi.org/10.1038/nphoton.2013.343} {\bibfield  {journal} {\bibinfo  {journal} {Nature Photonics}\ }\textbf {\bibinfo {volume} {8}},\ \bibinfo {pages} {145} (\bibinfo {year} {2014})}\BibitemShut {NoStop}%
\bibitem [{\citenamefont {Brasch}\ \emph {et~al.}(2016)\citenamefont {Brasch}, \citenamefont {Geiselmann}, \citenamefont {Herr}, \citenamefont {Lihachev}, \citenamefont {Pfeiffer}, \citenamefont {Gorodetsky},\ and\ \citenamefont {Kippenberg}}]{brasch_photonic_2016}%
  \BibitemOpen
  \bibfield  {author} {\bibinfo {author} {\bibfnamefont {V.}~\bibnamefont {Brasch}}, \bibinfo {author} {\bibfnamefont {M.}~\bibnamefont {Geiselmann}}, \bibinfo {author} {\bibfnamefont {T.}~\bibnamefont {Herr}}, \bibinfo {author} {\bibfnamefont {G.}~\bibnamefont {Lihachev}}, \bibinfo {author} {\bibfnamefont {M.~H.~P.}\ \bibnamefont {Pfeiffer}}, \bibinfo {author} {\bibfnamefont {M.~L.}\ \bibnamefont {Gorodetsky}},\ and\ \bibinfo {author} {\bibfnamefont {T.~J.}\ \bibnamefont {Kippenberg}},\ }\bibfield  {title} {\bibinfo {title} {Photonic chip–based optical frequency comb using soliton {Cherenkov} radiation},\ }\href {https://doi.org/10.1126/science.aad4811} {\bibfield  {journal} {\bibinfo  {journal} {Science}\ }\textbf {\bibinfo {volume} {351}},\ \bibinfo {pages} {357} (\bibinfo {year} {2016})}\BibitemShut {NoStop}%
\bibitem [{\citenamefont {Englebert}\ \emph {et~al.}(2023)\citenamefont {Englebert}, \citenamefont {Arabí}, \citenamefont {Gorza},\ and\ \citenamefont {Leo}}]{englebert_high_2023}%
  \BibitemOpen
  \bibfield  {author} {\bibinfo {author} {\bibfnamefont {N.}~\bibnamefont {Englebert}}, \bibinfo {author} {\bibfnamefont {C.~M.}\ \bibnamefont {Arabí}}, \bibinfo {author} {\bibfnamefont {S.-P.}\ \bibnamefont {Gorza}},\ and\ \bibinfo {author} {\bibfnamefont {F.}~\bibnamefont {Leo}},\ }\bibfield  {title} {\bibinfo {title} {High peak-to-background-ratio solitons in a coherently driven active fiber cavity},\ }\href {https://doi.org/10.1063/5.0159693} {\bibfield  {journal} {\bibinfo  {journal} {APL Photonics}\ }\textbf {\bibinfo {volume} {8}},\ \bibinfo {pages} {120802} (\bibinfo {year} {2023})}\BibitemShut {NoStop}%
\bibitem [{\citenamefont {Li}\ \emph {et~al.}(2020)\citenamefont {Li}, \citenamefont {Xu}, \citenamefont {Coen}, \citenamefont {Murdoch},\ and\ \citenamefont {Erkintalo}}]{li_experimental_2020}%
  \BibitemOpen
  \bibfield  {author} {\bibinfo {author} {\bibfnamefont {Z.}~\bibnamefont {Li}}, \bibinfo {author} {\bibfnamefont {Y.}~\bibnamefont {Xu}}, \bibinfo {author} {\bibfnamefont {S.}~\bibnamefont {Coen}}, \bibinfo {author} {\bibfnamefont {S.~G.}\ \bibnamefont {Murdoch}},\ and\ \bibinfo {author} {\bibfnamefont {M.}~\bibnamefont {Erkintalo}},\ }\bibfield  {title} {\bibinfo {title} {Experimental observations of bright dissipative cavity solitons and their collapsed snaking in a {Kerr} resonator with normal dispersion driving},\ }\href {https://doi.org/10.1364/OPTICA.400646} {\bibfield  {journal} {\bibinfo  {journal} {Optica}\ }\textbf {\bibinfo {volume} {7}},\ \bibinfo {pages} {1195} (\bibinfo {year} {2020})}\BibitemShut {NoStop}%
\bibitem [{\citenamefont {Englebert}\ \emph {et~al.}(2021)\citenamefont {Englebert}, \citenamefont {Mas~Arabí}, \citenamefont {Parra-Rivas}, \citenamefont {Gorza},\ and\ \citenamefont {Leo}}]{englebert_temporal_2021}%
  \BibitemOpen
  \bibfield  {author} {\bibinfo {author} {\bibfnamefont {N.}~\bibnamefont {Englebert}}, \bibinfo {author} {\bibfnamefont {C.}~\bibnamefont {Mas~Arabí}}, \bibinfo {author} {\bibfnamefont {P.}~\bibnamefont {Parra-Rivas}}, \bibinfo {author} {\bibfnamefont {S.-P.}\ \bibnamefont {Gorza}},\ and\ \bibinfo {author} {\bibfnamefont {F.}~\bibnamefont {Leo}},\ }\bibfield  {title} {\bibinfo {title} {Temporal solitons in a coherently driven active resonator},\ }\href {https://doi.org/10.1038/s41566-021-00807-w} {\bibfield  {journal} {\bibinfo  {journal} {Nature Photonics}\ }\textbf {\bibinfo {volume} {15}},\ \bibinfo {pages} {536} (\bibinfo {year} {2021})}\BibitemShut {NoStop}%
\bibitem [{\citenamefont {Obrzud}\ \emph {et~al.}(2017)\citenamefont {Obrzud}, \citenamefont {Lecomte},\ and\ \citenamefont {Herr}}]{obrzud_temporal_2017}%
  \BibitemOpen
  \bibfield  {author} {\bibinfo {author} {\bibfnamefont {E.}~\bibnamefont {Obrzud}}, \bibinfo {author} {\bibfnamefont {S.}~\bibnamefont {Lecomte}},\ and\ \bibinfo {author} {\bibfnamefont {T.}~\bibnamefont {Herr}},\ }\bibfield  {title} {\bibinfo {title} {Temporal solitons in microresonators driven by optical pulses},\ }\href {https://doi.org/10.1038/nphoton.2017.140} {\bibfield  {journal} {\bibinfo  {journal} {Nature Photonics}\ }\textbf {\bibinfo {volume} {11}},\ \bibinfo {pages} {600} (\bibinfo {year} {2017})}\BibitemShut {NoStop}%
\bibitem [{\citenamefont {Bunel}\ \emph {et~al.}(2024{\natexlab{a}})\citenamefont {Bunel}, \citenamefont {Conforti}, \citenamefont {Ziani}, \citenamefont {Lumeau}, \citenamefont {Moreau}, \citenamefont {Fernandez}, \citenamefont {Llopis}, \citenamefont {Bourcier},\ and\ \citenamefont {Mussot}}]{bunel_28_2024}%
  \BibitemOpen
  \bibfield  {author} {\bibinfo {author} {\bibfnamefont {T.}~\bibnamefont {Bunel}}, \bibinfo {author} {\bibfnamefont {M.}~\bibnamefont {Conforti}}, \bibinfo {author} {\bibfnamefont {Z.}~\bibnamefont {Ziani}}, \bibinfo {author} {\bibfnamefont {J.}~\bibnamefont {Lumeau}}, \bibinfo {author} {\bibfnamefont {A.}~\bibnamefont {Moreau}}, \bibinfo {author} {\bibfnamefont {A.}~\bibnamefont {Fernandez}}, \bibinfo {author} {\bibfnamefont {O.}~\bibnamefont {Llopis}}, \bibinfo {author} {\bibfnamefont {G.}~\bibnamefont {Bourcier}},\ and\ \bibinfo {author} {\bibfnamefont {A.}~\bibnamefont {Mussot}},\ }\bibfield  {title} {\bibinfo {title} {28 {THz} soliton frequency comb in a continuous-wave pumped fiber {Fabry}–{Pérot} resonator},\ }\href {https://doi.org/10.1063/5.0176533} {\bibfield  {journal} {\bibinfo  {journal} {APL Photonics}\ }\textbf {\bibinfo {volume} {9}},\ \bibinfo {pages} {010804} (\bibinfo {year} {2024}{\natexlab{a}})}\BibitemShut {NoStop}%
\bibitem [{\citenamefont {Lucas}\ \emph {et~al.}(2023)\citenamefont {Lucas}, \citenamefont {Deroh},\ and\ \citenamefont {Kibler}}]{lucas_dynamic_2023}%
  \BibitemOpen
  \bibfield  {author} {\bibinfo {author} {\bibfnamefont {E.}~\bibnamefont {Lucas}}, \bibinfo {author} {\bibfnamefont {M.}~\bibnamefont {Deroh}},\ and\ \bibinfo {author} {\bibfnamefont {B.}~\bibnamefont {Kibler}},\ }\bibfield  {title} {\bibinfo {title} {Dynamic {Interplay} {Between} {Kerr} {Combs} and {Brillouin} {Lasing} in {Fiber} {Cavities}},\ }\href {https://doi.org/10.1002/lpor.202300041} {\bibfield  {journal} {\bibinfo  {journal} {Laser \& Photonics Reviews}\ ,\ \bibinfo {pages} {2300041}} (\bibinfo {year} {2023})}\BibitemShut {NoStop}%
\bibitem [{\citenamefont {Nie}\ \emph {et~al.}(2022)\citenamefont {Nie}, \citenamefont {Li}, \citenamefont {Jia}, \citenamefont {Xie}, \citenamefont {Yan}, \citenamefont {Zhu}, \citenamefont {Xie},\ and\ \citenamefont {Huang}}]{nie_dissipative_2022}%
  \BibitemOpen
  \bibfield  {author} {\bibinfo {author} {\bibfnamefont {M.}~\bibnamefont {Nie}}, \bibinfo {author} {\bibfnamefont {B.}~\bibnamefont {Li}}, \bibinfo {author} {\bibfnamefont {K.}~\bibnamefont {Jia}}, \bibinfo {author} {\bibfnamefont {Y.}~\bibnamefont {Xie}}, \bibinfo {author} {\bibfnamefont {J.}~\bibnamefont {Yan}}, \bibinfo {author} {\bibfnamefont {S.}~\bibnamefont {Zhu}}, \bibinfo {author} {\bibfnamefont {Z.}~\bibnamefont {Xie}},\ and\ \bibinfo {author} {\bibfnamefont {S.-W.}\ \bibnamefont {Huang}},\ }\bibfield  {title} {\bibinfo {title} {Dissipative soliton generation and real-time dynamics in microresonator-filtered fiber lasers},\ }\href {https://doi.org/10.1038/s41377-022-00998-z} {\bibfield  {journal} {\bibinfo  {journal} {Light: Science \& Applications}\ }\textbf {\bibinfo {volume} {11}},\ \bibinfo {pages} {296} (\bibinfo {year} {2022})}\BibitemShut {NoStop}%
\bibitem [{\citenamefont {Bunel}\ \emph {et~al.}(2025)\citenamefont {Bunel}, \citenamefont {Lumeau}, \citenamefont {Moreau}, \citenamefont {Fernandez}, \citenamefont {Llopis}, \citenamefont {Bourcier}, \citenamefont {Perego}, \citenamefont {Conforti},\ and\ \citenamefont {Mussot}}]{bunel2025brillouininducedkerrfrequencycomb}%
  \BibitemOpen
  \bibfield  {author} {\bibinfo {author} {\bibfnamefont {T.}~\bibnamefont {Bunel}}, \bibinfo {author} {\bibfnamefont {J.}~\bibnamefont {Lumeau}}, \bibinfo {author} {\bibfnamefont {A.}~\bibnamefont {Moreau}}, \bibinfo {author} {\bibfnamefont {A.}~\bibnamefont {Fernandez}}, \bibinfo {author} {\bibfnamefont {O.}~\bibnamefont {Llopis}}, \bibinfo {author} {\bibfnamefont {G.}~\bibnamefont {Bourcier}}, \bibinfo {author} {\bibfnamefont {A.}~\bibnamefont {Perego}}, \bibinfo {author} {\bibfnamefont {M.}~\bibnamefont {Conforti}},\ and\ \bibinfo {author} {\bibfnamefont {A.}~\bibnamefont {Mussot}},\ }\href {https://arxiv.org/abs/2502.03037} {\bibinfo {title} {Brillouin-induced kerr frequency comb in normal dispersion fiber fabry perot resonators}} (\bibinfo {year} {2025}),\ \Eprint {https://arxiv.org/abs/2502.03037} {arXiv:2502.03037 [physics.optics]} \BibitemShut {NoStop}%
\bibitem [{\citenamefont {Li}\ \emph {et~al.}(2025)\citenamefont {Li}, \citenamefont {Chen},\ and\ \citenamefont {Wu}}]{Li_Chen_Wu_2025}%
  \BibitemOpen
  \bibfield  {author} {\bibinfo {author} {\bibfnamefont {T.}~\bibnamefont {Li}}, \bibinfo {author} {\bibfnamefont {J.}~\bibnamefont {Chen}},\ and\ \bibinfo {author} {\bibfnamefont {K.}~\bibnamefont {Wu}},\ }\bibfield  {title} {\bibinfo {title} {Ultra‐flat broadband low‐noise frequency comb in a fiber fabry‐perot resonator},\ }\bibfield  {journal} {\bibinfo  {journal} {Laser and Photonics Reviews}\ }\href {https://doi.org/10.1002/lpor.202400180} {10.1002/lpor.202400180} (\bibinfo {year} {2025})\BibitemShut {NoStop}%
\bibitem [{\citenamefont {Li}\ \emph {et~al.}(2023{\natexlab{a}})\citenamefont {Li}, \citenamefont {Wu}, \citenamefont {Zhang}, \citenamefont {Cai},\ and\ \citenamefont {Chen}}]{li_experimental_2023}%
  \BibitemOpen
  \bibfield  {author} {\bibinfo {author} {\bibfnamefont {T.}~\bibnamefont {Li}}, \bibinfo {author} {\bibfnamefont {K.}~\bibnamefont {Wu}}, \bibinfo {author} {\bibfnamefont {X.}~\bibnamefont {Zhang}}, \bibinfo {author} {\bibfnamefont {M.}~\bibnamefont {Cai}},\ and\ \bibinfo {author} {\bibfnamefont {J.}~\bibnamefont {Chen}},\ }\bibfield  {title} {\bibinfo {title} {Experimental observation of stimulated {Raman} scattering enabled localized structure in a normal dispersion {FP} resonator},\ }\href {https://doi.org/10.1364/OPTICA.496225} {\bibfield  {journal} {\bibinfo  {journal} {Optica}\ }\textbf {\bibinfo {volume} {10}},\ \bibinfo {pages} {1389} (\bibinfo {year} {2023}{\natexlab{a}})}\BibitemShut {NoStop}%
\bibitem [{\citenamefont {Li}\ \emph {et~al.}(2023{\natexlab{b}})\citenamefont {Li}, \citenamefont {Xu}, \citenamefont {Shamailov}, \citenamefont {Wen}, \citenamefont {Wang}, \citenamefont {Wei}, \citenamefont {Yang}, \citenamefont {Coen}, \citenamefont {Murdoch},\ and\ \citenamefont {Erkintalo}}]{li_ultrashort_2023}%
  \BibitemOpen
  \bibfield  {author} {\bibinfo {author} {\bibfnamefont {Z.}~\bibnamefont {Li}}, \bibinfo {author} {\bibfnamefont {Y.}~\bibnamefont {Xu}}, \bibinfo {author} {\bibfnamefont {S.}~\bibnamefont {Shamailov}}, \bibinfo {author} {\bibfnamefont {X.}~\bibnamefont {Wen}}, \bibinfo {author} {\bibfnamefont {W.}~\bibnamefont {Wang}}, \bibinfo {author} {\bibfnamefont {X.}~\bibnamefont {Wei}}, \bibinfo {author} {\bibfnamefont {Z.}~\bibnamefont {Yang}}, \bibinfo {author} {\bibfnamefont {S.}~\bibnamefont {Coen}}, \bibinfo {author} {\bibfnamefont {S.~G.}\ \bibnamefont {Murdoch}},\ and\ \bibinfo {author} {\bibfnamefont {M.}~\bibnamefont {Erkintalo}},\ }\bibfield  {title} {\bibinfo {title} {Ultrashort dissipative {Raman} solitons in {Kerr} resonators driven with phase-coherent optical pulses},\ }\bibfield  {journal} {\bibinfo  {journal} {Nature Photonics}\ }\href {https://doi.org/10.1038/s41566-023-01303-z} {10.1038/s41566-023-01303-z} (\bibinfo {year} {2023}{\natexlab{b}})\BibitemShut {NoStop}%
\bibitem [{\citenamefont {Suh}\ and\ \citenamefont {Vahala}(2018)}]{suh_soliton_2018}%
  \BibitemOpen
  \bibfield  {author} {\bibinfo {author} {\bibfnamefont {M.-G.}\ \bibnamefont {Suh}}\ and\ \bibinfo {author} {\bibfnamefont {K.~J.}\ \bibnamefont {Vahala}},\ }\bibfield  {title} {\bibinfo {title} {Soliton microcomb range measurement},\ }\href {https://doi.org/10.1126/science.aao1968} {\bibfield  {journal} {\bibinfo  {journal} {Science}\ }\textbf {\bibinfo {volume} {359}},\ \bibinfo {pages} {884} (\bibinfo {year} {2018})}\BibitemShut {NoStop}%
\bibitem [{\citenamefont {Yang}\ \emph {et~al.}(2016)\citenamefont {Yang}, \citenamefont {Yi}, \citenamefont {Yang},\ and\ \citenamefont {Vahala}}]{Yang_Yi_Yang_Vahala_2016}%
  \BibitemOpen
  \bibfield  {author} {\bibinfo {author} {\bibfnamefont {Q.-F.}\ \bibnamefont {Yang}}, \bibinfo {author} {\bibfnamefont {X.}~\bibnamefont {Yi}}, \bibinfo {author} {\bibfnamefont {K.~Y.}\ \bibnamefont {Yang}},\ and\ \bibinfo {author} {\bibfnamefont {K.}~\bibnamefont {Vahala}},\ }\bibfield  {title} {\bibinfo {title} {Stokes solitons in optical microcavities},\ }\href {https://doi.org/10.1038/nphys3875} {\bibfield  {journal} {\bibinfo  {journal} {Nature Physics}\ }\textbf {\bibinfo {volume} {13}},\ \bibinfo {pages} {53–57} (\bibinfo {year} {2016})}\BibitemShut {NoStop}%
\bibitem [{\citenamefont {Wang}\ \emph {et~al.}(2018)\citenamefont {Wang}, \citenamefont {Anderson}, \citenamefont {Coen}, \citenamefont {Murdoch},\ and\ \citenamefont {Erkintalo}}]{wang_stimulated_2018}%
  \BibitemOpen
  \bibfield  {author} {\bibinfo {author} {\bibfnamefont {Y.}~\bibnamefont {Wang}}, \bibinfo {author} {\bibfnamefont {M.}~\bibnamefont {Anderson}}, \bibinfo {author} {\bibfnamefont {S.}~\bibnamefont {Coen}}, \bibinfo {author} {\bibfnamefont {S.~G.}\ \bibnamefont {Murdoch}},\ and\ \bibinfo {author} {\bibfnamefont {M.}~\bibnamefont {Erkintalo}},\ }\bibfield  {title} {\bibinfo {title} {Stimulated {Raman} {Scattering} {Imposes} {Fundamental} {Limits} to the {Duration} and {Bandwidth} of {Temporal} {Cavity} {Solitons}},\ }\bibfield  {journal} {\bibinfo  {journal} {Physical Review Letters}\ }\textbf {\bibinfo {volume} {120}},\ \href {https://doi.org/10.1103/physrevlett.120.053902} {10.1103/physrevlett.120.053902} (\bibinfo {year} {2018}),\ \bibinfo {note} {publisher: American Physical Society (APS)}\BibitemShut {NoStop}%
\bibitem [{\citenamefont {Milián}\ \emph {et~al.}(2015)\citenamefont {Milián}, \citenamefont {Gorbach}, \citenamefont {Taki}, \citenamefont {Yulin},\ and\ \citenamefont {Skryabin}}]{milian_solitons_2015}%
  \BibitemOpen
  \bibfield  {author} {\bibinfo {author} {\bibfnamefont {C.}~\bibnamefont {Milián}}, \bibinfo {author} {\bibfnamefont {A.~V.}\ \bibnamefont {Gorbach}}, \bibinfo {author} {\bibfnamefont {M.}~\bibnamefont {Taki}}, \bibinfo {author} {\bibfnamefont {A.~V.}\ \bibnamefont {Yulin}},\ and\ \bibinfo {author} {\bibfnamefont {D.~V.}\ \bibnamefont {Skryabin}},\ }\bibfield  {title} {\bibinfo {title} {Solitons and frequency combs in silica microring resonators: {Interplay} of the {Raman} and higher-order dispersion effects},\ }\href {https://doi.org/10.1103/PhysRevA.92.033851} {\bibfield  {journal} {\bibinfo  {journal} {Physical Review A}\ }\textbf {\bibinfo {volume} {92}},\ \bibinfo {pages} {033851} (\bibinfo {year} {2015})}\BibitemShut {NoStop}%
\bibitem [{\citenamefont {Karpov}\ \emph {et~al.}(2016)\citenamefont {Karpov}, \citenamefont {Guo}, \citenamefont {Kordts}, \citenamefont {Brasch}, \citenamefont {Pfeiffer}, \citenamefont {Zervas}, \citenamefont {Geiselmann},\ and\ \citenamefont {Kippenberg}}]{karpov_raman_2016}%
  \BibitemOpen
  \bibfield  {author} {\bibinfo {author} {\bibfnamefont {M.}~\bibnamefont {Karpov}}, \bibinfo {author} {\bibfnamefont {H.}~\bibnamefont {Guo}}, \bibinfo {author} {\bibfnamefont {A.}~\bibnamefont {Kordts}}, \bibinfo {author} {\bibfnamefont {V.}~\bibnamefont {Brasch}}, \bibinfo {author} {\bibfnamefont {M.~H.}\ \bibnamefont {Pfeiffer}}, \bibinfo {author} {\bibfnamefont {M.}~\bibnamefont {Zervas}}, \bibinfo {author} {\bibfnamefont {M.}~\bibnamefont {Geiselmann}},\ and\ \bibinfo {author} {\bibfnamefont {T.~J.}\ \bibnamefont {Kippenberg}},\ }\bibfield  {title} {\bibinfo {title} {Raman {Self}-{Frequency} {Shift} of {Dissipative} {Kerr} {Solitons} in an {Optical} {Microresonator}},\ }\bibfield  {journal} {\bibinfo  {journal} {Physical Review Letters}\ }\textbf {\bibinfo {volume} {116}},\ \href {https://doi.org/10.1103/physrevlett.116.103902} {10.1103/physrevlett.116.103902} (\bibinfo {year} {2016}),\ \bibinfo {note} {publisher: American Physical Society (APS)}\BibitemShut {NoStop}%
\bibitem [{\citenamefont {Dudley}\ and\ \citenamefont {Taylor}(2010)}]{Supercontinuum_2010}%
  \BibitemOpen
  \bibinfo {editor} {\bibfnamefont {J.~M.}\ \bibnamefont {Dudley}}\ and\ \bibinfo {editor} {\bibfnamefont {J.~R.}\ \bibnamefont {Taylor}},\ eds.,\ \href {https://doi.org/10.1017/cbo9780511750465} {\emph {\bibinfo {title} {Supercontinuum Generation in Optical Fibers}}}\ (\bibinfo  {publisher} {Cambridge University Press},\ \bibinfo {year} {2010})\BibitemShut {NoStop}%
\bibitem [{\citenamefont {Dudley}\ \emph {et~al.}(2006)\citenamefont {Dudley}, \citenamefont {Genty},\ and\ \citenamefont {Coen}}]{dudley_supercontinuum_2006}%
  \BibitemOpen
  \bibfield  {author} {\bibinfo {author} {\bibfnamefont {J.~M.}\ \bibnamefont {Dudley}}, \bibinfo {author} {\bibfnamefont {G.}~\bibnamefont {Genty}},\ and\ \bibinfo {author} {\bibfnamefont {S.}~\bibnamefont {Coen}},\ }\bibfield  {title} {\bibinfo {title} {Supercontinuum generation in photonic crystal fiber},\ }\href {https://doi.org/10.1103/RevModPhys.78.1135} {\bibfield  {journal} {\bibinfo  {journal} {Reviews of Modern Physics}\ }\textbf {\bibinfo {volume} {78}},\ \bibinfo {pages} {1135} (\bibinfo {year} {2006})}\BibitemShut {NoStop}%
\bibitem [{\citenamefont {Brès}\ \emph {et~al.}(2023)\citenamefont {Brès}, \citenamefont {Della~Torre}, \citenamefont {Grassani}, \citenamefont {Brasch}, \citenamefont {Grillet},\ and\ \citenamefont {Monat}}]{bres_supercontinuum_2023}%
  \BibitemOpen
  \bibfield  {author} {\bibinfo {author} {\bibfnamefont {C.-S.}\ \bibnamefont {Brès}}, \bibinfo {author} {\bibfnamefont {A.}~\bibnamefont {Della~Torre}}, \bibinfo {author} {\bibfnamefont {D.}~\bibnamefont {Grassani}}, \bibinfo {author} {\bibfnamefont {V.}~\bibnamefont {Brasch}}, \bibinfo {author} {\bibfnamefont {C.}~\bibnamefont {Grillet}},\ and\ \bibinfo {author} {\bibfnamefont {C.}~\bibnamefont {Monat}},\ }\bibfield  {title} {\bibinfo {title} {Supercontinuum in integrated photonics: generation, applications, challenges, and perspectives},\ }\href {https://doi.org/10.1515/nanoph-2022-0749} {\bibfield  {journal} {\bibinfo  {journal} {Nanophotonics}\ }\textbf {\bibinfo {volume} {12}},\ \bibinfo {pages} {1199} (\bibinfo {year} {2023})}\BibitemShut {NoStop}%
\bibitem [{\citenamefont {Meng}\ \emph {et~al.}(2021)\citenamefont {Meng}, \citenamefont {Lapre}, \citenamefont {Billet}, \citenamefont {Sylvestre}, \citenamefont {Merolla}, \citenamefont {Finot}, \citenamefont {Turitsyn}, \citenamefont {Genty},\ and\ \citenamefont {Dudley}}]{Meng_2021}%
  \BibitemOpen
  \bibfield  {author} {\bibinfo {author} {\bibfnamefont {F.}~\bibnamefont {Meng}}, \bibinfo {author} {\bibfnamefont {C.}~\bibnamefont {Lapre}}, \bibinfo {author} {\bibfnamefont {C.}~\bibnamefont {Billet}}, \bibinfo {author} {\bibfnamefont {T.}~\bibnamefont {Sylvestre}}, \bibinfo {author} {\bibfnamefont {J.-M.}\ \bibnamefont {Merolla}}, \bibinfo {author} {\bibfnamefont {C.}~\bibnamefont {Finot}}, \bibinfo {author} {\bibfnamefont {S.~K.}\ \bibnamefont {Turitsyn}}, \bibinfo {author} {\bibfnamefont {G.}~\bibnamefont {Genty}},\ and\ \bibinfo {author} {\bibfnamefont {J.~M.}\ \bibnamefont {Dudley}},\ }\bibfield  {title} {\bibinfo {title} {Intracavity incoherent supercontinuum dynamics and rogue waves in a broadband dissipative soliton laser},\ }\bibfield  {journal} {\bibinfo  {journal} {Nature Communications}\ }\textbf {\bibinfo {volume} {12}},\ \href {https://doi.org/10.1038/s41467-021-25861-4} {10.1038/s41467-021-25861-4} (\bibinfo {year} {2021})\BibitemShut {NoStop}%
\bibitem [{\citenamefont {Del’Haye}\ \emph {et~al.}(2007)\citenamefont {Del’Haye}, \citenamefont {Schliesser}, \citenamefont {Arcizet}, \citenamefont {Wilken}, \citenamefont {Holzwarth},\ and\ \citenamefont {Kippenberg}}]{delhaye_optical_2007}%
  \BibitemOpen
  \bibfield  {author} {\bibinfo {author} {\bibfnamefont {P.}~\bibnamefont {Del’Haye}}, \bibinfo {author} {\bibfnamefont {A.}~\bibnamefont {Schliesser}}, \bibinfo {author} {\bibfnamefont {O.}~\bibnamefont {Arcizet}}, \bibinfo {author} {\bibfnamefont {T.}~\bibnamefont {Wilken}}, \bibinfo {author} {\bibfnamefont {R.}~\bibnamefont {Holzwarth}},\ and\ \bibinfo {author} {\bibfnamefont {T.~J.}\ \bibnamefont {Kippenberg}},\ }\bibfield  {title} {\bibinfo {title} {Optical frequency comb generation from a monolithic microresonator},\ }\href {https://doi.org/10.1038/nature06401} {\bibfield  {journal} {\bibinfo  {journal} {Nature}\ }\textbf {\bibinfo {volume} {450}},\ \bibinfo {pages} {1214} (\bibinfo {year} {2007})}\BibitemShut {NoStop}%
\bibitem [{\citenamefont {Papp}\ \emph {et~al.}(2014)\citenamefont {Papp}, \citenamefont {Beha}, \citenamefont {Del’Haye}, \citenamefont {Quinlan}, \citenamefont {Lee}, \citenamefont {Vahala},\ and\ \citenamefont {Diddams}}]{papp_microresonator_2014}%
  \BibitemOpen
  \bibfield  {author} {\bibinfo {author} {\bibfnamefont {S.~B.}\ \bibnamefont {Papp}}, \bibinfo {author} {\bibfnamefont {K.}~\bibnamefont {Beha}}, \bibinfo {author} {\bibfnamefont {P.}~\bibnamefont {Del’Haye}}, \bibinfo {author} {\bibfnamefont {F.}~\bibnamefont {Quinlan}}, \bibinfo {author} {\bibfnamefont {H.}~\bibnamefont {Lee}}, \bibinfo {author} {\bibfnamefont {K.~J.}\ \bibnamefont {Vahala}},\ and\ \bibinfo {author} {\bibfnamefont {S.~A.}\ \bibnamefont {Diddams}},\ }\bibfield  {title} {\bibinfo {title} {Microresonator frequency comb optical clock},\ }\href {https://doi.org/10.1364/optica.1.000010} {\bibfield  {journal} {\bibinfo  {journal} {Optica}\ }\textbf {\bibinfo {volume} {1}},\ \bibinfo {pages} {10} (\bibinfo {year} {2014})},\ \bibinfo {note} {publisher: Optica Publishing Group}\BibitemShut {NoStop}%
\bibitem [{\citenamefont {Li}\ \emph {et~al.}(2012)\citenamefont {Li}, \citenamefont {Lee}, \citenamefont {Chen},\ and\ \citenamefont {Vahala}}]{li_low-pump-power_2012}%
  \BibitemOpen
  \bibfield  {author} {\bibinfo {author} {\bibfnamefont {J.}~\bibnamefont {Li}}, \bibinfo {author} {\bibfnamefont {H.}~\bibnamefont {Lee}}, \bibinfo {author} {\bibfnamefont {T.}~\bibnamefont {Chen}},\ and\ \bibinfo {author} {\bibfnamefont {K.~J.}\ \bibnamefont {Vahala}},\ }\bibfield  {title} {\bibinfo {title} {Low-{Pump}-{Power}, {Low}-{Phase}-{Noise}, and {Microwave} to {Millimeter}-{Wave} {Repetition} {Rate} {Operation} in {Microcombs}},\ }\bibfield  {journal} {\bibinfo  {journal} {Physical Review Letters}\ }\textbf {\bibinfo {volume} {109}},\ \href {https://doi.org/10.1103/physrevlett.109.233901} {10.1103/physrevlett.109.233901} (\bibinfo {year} {2012}),\ \bibinfo {note} {publisher: American Physical Society (APS)}\BibitemShut {NoStop}%
\bibitem [{\citenamefont {Xiao}\ \emph {et~al.}(2023)\citenamefont {Xiao}, \citenamefont {Li}, \citenamefont {Cai}, \citenamefont {Zhang}, \citenamefont {Huang}, \citenamefont {Li}, \citenamefont {Yao}, \citenamefont {Wu},\ and\ \citenamefont {Chen}}]{xiao_near-zero-dispersion_2023}%
  \BibitemOpen
  \bibfield  {author} {\bibinfo {author} {\bibfnamefont {Z.}~\bibnamefont {Xiao}}, \bibinfo {author} {\bibfnamefont {T.}~\bibnamefont {Li}}, \bibinfo {author} {\bibfnamefont {M.}~\bibnamefont {Cai}}, \bibinfo {author} {\bibfnamefont {H.}~\bibnamefont {Zhang}}, \bibinfo {author} {\bibfnamefont {Y.}~\bibnamefont {Huang}}, \bibinfo {author} {\bibfnamefont {C.}~\bibnamefont {Li}}, \bibinfo {author} {\bibfnamefont {B.}~\bibnamefont {Yao}}, \bibinfo {author} {\bibfnamefont {K.}~\bibnamefont {Wu}},\ and\ \bibinfo {author} {\bibfnamefont {J.}~\bibnamefont {Chen}},\ }\bibfield  {title} {\bibinfo {title} {Near-zero-dispersion soliton and broadband modulational instability {Kerr} microcombs in anomalous dispersion},\ }\href {https://doi.org/10.1038/s41377-023-01076-8} {\bibfield  {journal} {\bibinfo  {journal} {Light: Science \& Applications}\ }\textbf {\bibinfo {volume} {12}},\ \bibinfo {pages} {33} (\bibinfo {year} {2023})}\BibitemShut {NoStop}%
\bibitem [{\citenamefont {Agrawal}(2013)}]{govind_p_agrawal_non_2013}%
  \BibitemOpen
  \bibfield  {author} {\bibinfo {author} {\bibfnamefont {G.~P.}\ \bibnamefont {Agrawal}},\ }\bibfield  {title} {\bibinfo {title} {Non linear fiber optics},\ }in\ \href {https://doi.org/10.1016/B978-0-12-397023-7.00018-8} {\emph {\bibinfo {booktitle} {Nonlinear {Fiber} {Optics}}}}\ (\bibinfo  {publisher} {Elsevier},\ \bibinfo {year} {2013})\ pp.\ \bibinfo {pages} {i--ii}\BibitemShut {NoStop}%
\bibitem [{\citenamefont {Zideluns}\ \emph {et~al.}(2021)\citenamefont {Zideluns}, \citenamefont {Lemarchand}, \citenamefont {Arhilger}, \citenamefont {Hagedorn},\ and\ \citenamefont {Lumeau}}]{zideluns_automated_2021}%
  \BibitemOpen
  \bibfield  {author} {\bibinfo {author} {\bibfnamefont {J.}~\bibnamefont {Zideluns}}, \bibinfo {author} {\bibfnamefont {F.}~\bibnamefont {Lemarchand}}, \bibinfo {author} {\bibfnamefont {D.}~\bibnamefont {Arhilger}}, \bibinfo {author} {\bibfnamefont {H.}~\bibnamefont {Hagedorn}},\ and\ \bibinfo {author} {\bibfnamefont {J.}~\bibnamefont {Lumeau}},\ }\bibfield  {title} {\bibinfo {title} {Automated optical monitoring wavelength selection for thin-film filters},\ }\href {https://doi.org/10.1364/OE.439033} {\bibfield  {journal} {\bibinfo  {journal} {Optics Express}\ }\textbf {\bibinfo {volume} {29}},\ \bibinfo {pages} {33398} (\bibinfo {year} {2021})}\BibitemShut {NoStop}%
\bibitem [{\citenamefont {Bunel}\ \emph {et~al.}(2024{\natexlab{b}})\citenamefont {Bunel}, \citenamefont {Conforti}, \citenamefont {Lumeau}, \citenamefont {Moreau},\ and\ \citenamefont {Mussot}}]{bunel_broadband_2024}%
  \BibitemOpen
  \bibfield  {author} {\bibinfo {author} {\bibfnamefont {T.}~\bibnamefont {Bunel}}, \bibinfo {author} {\bibfnamefont {M.}~\bibnamefont {Conforti}}, \bibinfo {author} {\bibfnamefont {J.}~\bibnamefont {Lumeau}}, \bibinfo {author} {\bibfnamefont {A.}~\bibnamefont {Moreau}},\ and\ \bibinfo {author} {\bibfnamefont {A.}~\bibnamefont {Mussot}},\ }\bibfield  {title} {\bibinfo {title} {Broadband kerr frequency comb in fiber fabry-perot resonators induced by switching waves},\ }\href {https://doi.org/10.1103/PhysRevA.109.063521} {\bibfield  {journal} {\bibinfo  {journal} {Phys. Rev. A}\ }\textbf {\bibinfo {volume} {109}},\ \bibinfo {pages} {063521} (\bibinfo {year} {2024}{\natexlab{b}})}\BibitemShut {NoStop}%
\bibitem [{\citenamefont {Bunel}\ \emph {et~al.}(2023)\citenamefont {Bunel}, \citenamefont {Ziani}, \citenamefont {Conforti}, \citenamefont {Lumeau}, \citenamefont {Moreau}, \citenamefont {Fernandez}, \citenamefont {Llopis}, \citenamefont {Bourcier}, \citenamefont {Perego},\ and\ \citenamefont {Mussot}}]{bunel_impact_2023}%
  \BibitemOpen
  \bibfield  {author} {\bibinfo {author} {\bibfnamefont {T.}~\bibnamefont {Bunel}}, \bibinfo {author} {\bibfnamefont {Z.}~\bibnamefont {Ziani}}, \bibinfo {author} {\bibfnamefont {M.}~\bibnamefont {Conforti}}, \bibinfo {author} {\bibfnamefont {J.}~\bibnamefont {Lumeau}}, \bibinfo {author} {\bibfnamefont {A.}~\bibnamefont {Moreau}}, \bibinfo {author} {\bibfnamefont {A.}~\bibnamefont {Fernandez}}, \bibinfo {author} {\bibfnamefont {O.}~\bibnamefont {Llopis}}, \bibinfo {author} {\bibfnamefont {G.}~\bibnamefont {Bourcier}}, \bibinfo {author} {\bibfnamefont {A.~M.}\ \bibnamefont {Perego}},\ and\ \bibinfo {author} {\bibfnamefont {A.}~\bibnamefont {Mussot}},\ }\bibfield  {title} {\bibinfo {title} {Impact of pump pulse duration on modulation instability {Kerr} frequency combs in fiber {Fabry}–{Pérot} resonators},\ }\href {https://doi.org/10.1364/OL.506100} {\bibfield  {journal} {\bibinfo  {journal} {Optics Letters}\ }\textbf {\bibinfo {volume} {48}},\ \bibinfo {pages} {5955} (\bibinfo {year} {2023})}\BibitemShut
  {NoStop}%
\bibitem [{\citenamefont {Black}(2001)}]{black_introduction_2001}%
  \BibitemOpen
  \bibfield  {author} {\bibinfo {author} {\bibfnamefont {E.~D.}\ \bibnamefont {Black}},\ }\bibfield  {title} {\bibinfo {title} {An introduction to {Pound}–{Drever}–{Hall} laser frequency stabilization},\ }\href {https://doi.org/10.1119/1.1286663} {\bibfield  {journal} {\bibinfo  {journal} {American Journal of Physics}\ }\textbf {\bibinfo {volume} {69}},\ \bibinfo {pages} {79} (\bibinfo {year} {2001})}\BibitemShut {NoStop}%
\bibitem [{\citenamefont {Firth}\ \emph {et~al.}(2021)\citenamefont {Firth}, \citenamefont {Geddes}, \citenamefont {Karst},\ and\ \citenamefont {Oppo}}]{firth_analytic_2021}%
  \BibitemOpen
  \bibfield  {author} {\bibinfo {author} {\bibfnamefont {W.~J.}\ \bibnamefont {Firth}}, \bibinfo {author} {\bibfnamefont {J.~B.}\ \bibnamefont {Geddes}}, \bibinfo {author} {\bibfnamefont {N.~J.}\ \bibnamefont {Karst}},\ and\ \bibinfo {author} {\bibfnamefont {G.-L.}\ \bibnamefont {Oppo}},\ }\bibfield  {title} {\bibinfo {title} {Analytic instability thresholds in folded kerr resonators of arbitrary finesse},\ }\href@noop {} {\bibfield  {journal} {\bibinfo  {journal} {Physical Review A}\ }\textbf {\bibinfo {volume} {103}},\ \bibinfo {pages} {023510} (\bibinfo {year} {2021})}\BibitemShut {NoStop}%
\bibitem [{\citenamefont {Cole}\ \emph {et~al.}(2018)\citenamefont {Cole}, \citenamefont {Gatti}, \citenamefont {Papp}, \citenamefont {Prati},\ and\ \citenamefont {Lugiato}}]{cole_theory_2018}%
  \BibitemOpen
  \bibfield  {author} {\bibinfo {author} {\bibfnamefont {D.~C.}\ \bibnamefont {Cole}}, \bibinfo {author} {\bibfnamefont {A.}~\bibnamefont {Gatti}}, \bibinfo {author} {\bibfnamefont {S.~B.}\ \bibnamefont {Papp}}, \bibinfo {author} {\bibfnamefont {F.}~\bibnamefont {Prati}},\ and\ \bibinfo {author} {\bibfnamefont {L.}~\bibnamefont {Lugiato}},\ }\bibfield  {title} {\bibinfo {title} {Theory of {Kerr} frequency combs in {Fabry}-{Perot} resonators},\ }\href {https://doi.org/10.1103/PhysRevA.98.013831} {\bibfield  {journal} {\bibinfo  {journal} {Physical Review A}\ }\textbf {\bibinfo {volume} {98}},\ \bibinfo {pages} {013831} (\bibinfo {year} {2018})}\BibitemShut {NoStop}%
\bibitem [{\citenamefont {Ziani}\ \emph {et~al.}(2024)\citenamefont {Ziani}, \citenamefont {Bunel}, \citenamefont {Perego}, \citenamefont {Mussot},\ and\ \citenamefont {Conforti}}]{ziani_theory_2024}%
  \BibitemOpen
  \bibfield  {author} {\bibinfo {author} {\bibfnamefont {Z.}~\bibnamefont {Ziani}}, \bibinfo {author} {\bibfnamefont {T.}~\bibnamefont {Bunel}}, \bibinfo {author} {\bibfnamefont {A.~M.}\ \bibnamefont {Perego}}, \bibinfo {author} {\bibfnamefont {A.}~\bibnamefont {Mussot}},\ and\ \bibinfo {author} {\bibfnamefont {M.}~\bibnamefont {Conforti}},\ }\bibfield  {title} {\bibinfo {title} {Theory of modulation instability in {Kerr} {Fabry}-{Perot} resonators beyond the mean-field limit},\ }\href {https://doi.org/10.1103/PhysRevA.109.013507} {\bibfield  {journal} {\bibinfo  {journal} {Physical Review A}\ }\textbf {\bibinfo {volume} {109}},\ \bibinfo {pages} {013507} (\bibinfo {year} {2024})}\BibitemShut {NoStop}%
\bibitem [{\citenamefont {Bessin}\ \emph {et~al.}(2017)\citenamefont {Bessin}, \citenamefont {Copie}, \citenamefont {Conforti}, \citenamefont {Kudlinski},\ and\ \citenamefont {Mussot}}]{bessin_modulation_2017}%
  \BibitemOpen
  \bibfield  {author} {\bibinfo {author} {\bibfnamefont {F.}~\bibnamefont {Bessin}}, \bibinfo {author} {\bibfnamefont {F.}~\bibnamefont {Copie}}, \bibinfo {author} {\bibfnamefont {M.}~\bibnamefont {Conforti}}, \bibinfo {author} {\bibfnamefont {A.}~\bibnamefont {Kudlinski}},\ and\ \bibinfo {author} {\bibfnamefont {A.}~\bibnamefont {Mussot}},\ }\bibfield  {title} {\bibinfo {title} {Modulation instability in the weak normal dispersion region of passive fiber ring cavities},\ }\href {https://doi.org/10.1364/OL.42.003730} {\bibfield  {journal} {\bibinfo  {journal} {Optics Letters}\ }\textbf {\bibinfo {volume} {42}},\ \bibinfo {pages} {3730} (\bibinfo {year} {2017})}\BibitemShut {NoStop}%
\bibitem [{\citenamefont {Sayson}\ \emph {et~al.}(2019)\citenamefont {Sayson}, \citenamefont {Bi}, \citenamefont {Ng}, \citenamefont {Pham}, \citenamefont {Trainor}, \citenamefont {Schwefel}, \citenamefont {Coen}, \citenamefont {Erkintalo},\ and\ \citenamefont {Murdoch}}]{sayson_octave-spanning_2019}%
  \BibitemOpen
  \bibfield  {author} {\bibinfo {author} {\bibfnamefont {N.~L.~B.}\ \bibnamefont {Sayson}}, \bibinfo {author} {\bibfnamefont {T.}~\bibnamefont {Bi}}, \bibinfo {author} {\bibfnamefont {V.}~\bibnamefont {Ng}}, \bibinfo {author} {\bibfnamefont {H.}~\bibnamefont {Pham}}, \bibinfo {author} {\bibfnamefont {L.~S.}\ \bibnamefont {Trainor}}, \bibinfo {author} {\bibfnamefont {H.~G.~L.}\ \bibnamefont {Schwefel}}, \bibinfo {author} {\bibfnamefont {S.}~\bibnamefont {Coen}}, \bibinfo {author} {\bibfnamefont {M.}~\bibnamefont {Erkintalo}},\ and\ \bibinfo {author} {\bibfnamefont {S.~G.}\ \bibnamefont {Murdoch}},\ }\bibfield  {title} {\bibinfo {title} {Octave-spanning tunable parametric oscillation in crystalline {Kerr} microresonators},\ }\href {https://doi.org/10.1038/s41566-019-0485-4} {\bibfield  {journal} {\bibinfo  {journal} {Nature Photonics}\ }\textbf {\bibinfo {volume} {13}},\ \bibinfo {pages} {701} (\bibinfo {year} {2019})}\BibitemShut {NoStop}%
\bibitem [{\citenamefont {Mussot}\ \emph {et~al.}(2008)\citenamefont {Mussot}, \citenamefont {Louvergneaux}, \citenamefont {Akhmediev}, \citenamefont {Reynaud}, \citenamefont {Delage},\ and\ \citenamefont {Taki}}]{mussot_optical_2008}%
  \BibitemOpen
  \bibfield  {author} {\bibinfo {author} {\bibfnamefont {A.}~\bibnamefont {Mussot}}, \bibinfo {author} {\bibfnamefont {E.}~\bibnamefont {Louvergneaux}}, \bibinfo {author} {\bibfnamefont {N.}~\bibnamefont {Akhmediev}}, \bibinfo {author} {\bibfnamefont {F.}~\bibnamefont {Reynaud}}, \bibinfo {author} {\bibfnamefont {L.}~\bibnamefont {Delage}},\ and\ \bibinfo {author} {\bibfnamefont {M.}~\bibnamefont {Taki}},\ }\bibfield  {title} {\bibinfo {title} {Optical {Fiber} {Systems} {Are} {Convectively} {Unstable}},\ }\href {https://doi.org/10.1103/PhysRevLett.101.113904} {\bibfield  {journal} {\bibinfo  {journal} {Physical Review Letters}\ }\textbf {\bibinfo {volume} {101}},\ \bibinfo {pages} {113904} (\bibinfo {year} {2008})}\BibitemShut {NoStop}%
\bibitem [{\citenamefont {Leo}\ \emph {et~al.}(2013)\citenamefont {Leo}, \citenamefont {Mussot}, \citenamefont {Kockaert}, \citenamefont {Emplit}, \citenamefont {Haelterman},\ and\ \citenamefont {Taki}}]{leo_nonlinear_2013}%
  \BibitemOpen
  \bibfield  {author} {\bibinfo {author} {\bibfnamefont {F.}~\bibnamefont {Leo}}, \bibinfo {author} {\bibfnamefont {A.}~\bibnamefont {Mussot}}, \bibinfo {author} {\bibfnamefont {P.}~\bibnamefont {Kockaert}}, \bibinfo {author} {\bibfnamefont {P.}~\bibnamefont {Emplit}}, \bibinfo {author} {\bibfnamefont {M.}~\bibnamefont {Haelterman}},\ and\ \bibinfo {author} {\bibfnamefont {M.}~\bibnamefont {Taki}},\ }\bibfield  {title} {\bibinfo {title} {Nonlinear {Symmetry} {Breaking} {Induced} by {Third}-{Order} {Dispersion} in {Optical} {Fiber} {Cavities}},\ }\href {https://doi.org/10.1103/PhysRevLett.110.104103} {\bibfield  {journal} {\bibinfo  {journal} {Physical Review Letters}\ }\textbf {\bibinfo {volume} {110}},\ \bibinfo {pages} {104103} (\bibinfo {year} {2013})}\BibitemShut {NoStop}%
\bibitem [{\citenamefont {Milián}\ and\ \citenamefont {Skryabin}(2014)}]{milian_soliton_2014}%
  \BibitemOpen
  \bibfield  {author} {\bibinfo {author} {\bibfnamefont {C.}~\bibnamefont {Milián}}\ and\ \bibinfo {author} {\bibfnamefont {D.}~\bibnamefont {Skryabin}},\ }\bibfield  {title} {\bibinfo {title} {Soliton families and resonant radiation in a micro-ring resonator near zero group-velocity dispersion},\ }\href@noop {} {\bibfield  {journal} {\bibinfo  {journal} {Optics Express}\ }\textbf {\bibinfo {volume} {22}},\ \bibinfo {pages} {3732} (\bibinfo {year} {2014})}\BibitemShut {NoStop}%
\bibitem [{\citenamefont {Macnaughtan}\ \emph {et~al.}(2023)\citenamefont {Macnaughtan}, \citenamefont {Erkintalo}, \citenamefont {Coen}, \citenamefont {Murdoch},\ and\ \citenamefont {Xu}}]{macnaughtan_temporal_2023}%
  \BibitemOpen
  \bibfield  {author} {\bibinfo {author} {\bibfnamefont {M.}~\bibnamefont {Macnaughtan}}, \bibinfo {author} {\bibfnamefont {M.}~\bibnamefont {Erkintalo}}, \bibinfo {author} {\bibfnamefont {S.}~\bibnamefont {Coen}}, \bibinfo {author} {\bibfnamefont {S.}~\bibnamefont {Murdoch}},\ and\ \bibinfo {author} {\bibfnamefont {Y.}~\bibnamefont {Xu}},\ }\bibfield  {title} {\bibinfo {title} {Temporal characteristics of stationary switching waves in a normal dispersion pulsed-pump fiber cavity},\ }\href {https://doi.org/10.1364/OL.492998} {\bibfield  {journal} {\bibinfo  {journal} {Optics Letters}\ }\textbf {\bibinfo {volume} {48}},\ \bibinfo {pages} {4097} (\bibinfo {year} {2023})}\BibitemShut {NoStop}%
\bibitem [{\citenamefont {Anderson}\ \emph {et~al.}(2022)\citenamefont {Anderson}, \citenamefont {Weng}, \citenamefont {Lihachev}, \citenamefont {Tikan}, \citenamefont {Liu},\ and\ \citenamefont {Kippenberg}}]{anderson_zero_2022}%
  \BibitemOpen
  \bibfield  {author} {\bibinfo {author} {\bibfnamefont {M.~H.}\ \bibnamefont {Anderson}}, \bibinfo {author} {\bibfnamefont {W.}~\bibnamefont {Weng}}, \bibinfo {author} {\bibfnamefont {G.}~\bibnamefont {Lihachev}}, \bibinfo {author} {\bibfnamefont {A.}~\bibnamefont {Tikan}}, \bibinfo {author} {\bibfnamefont {J.}~\bibnamefont {Liu}},\ and\ \bibinfo {author} {\bibfnamefont {T.~J.}\ \bibnamefont {Kippenberg}},\ }\bibfield  {title} {\bibinfo {title} {Zero dispersion {Kerr} solitons in optical microresonators},\ }\href {https://doi.org/10.1038/s41467-022-31916-x} {\bibfield  {journal} {\bibinfo  {journal} {Nature Communications}\ }\textbf {\bibinfo {volume} {13}},\ \bibinfo {pages} {4764} (\bibinfo {year} {2022})}\BibitemShut {NoStop}%
\bibitem [{\citenamefont {Zhang}\ \emph {et~al.}(2023)\citenamefont {Zhang}, \citenamefont {Bi},\ and\ \citenamefont {Del'Haye}}]{zhang_quintic_2023}%
  \BibitemOpen
  \bibfield  {author} {\bibinfo {author} {\bibfnamefont {S.}~\bibnamefont {Zhang}}, \bibinfo {author} {\bibfnamefont {T.}~\bibnamefont {Bi}},\ and\ \bibinfo {author} {\bibfnamefont {P.}~\bibnamefont {Del'Haye}},\ }\bibfield  {title} {\bibinfo {title} {Quintic {Dispersion} {Soliton} {Frequency} {Combs} in a {Microresonator}},\ }\href {https://doi.org/10.1002/lpor.202300075} {\bibfield  {journal} {\bibinfo  {journal} {Laser \& Photonics Reviews}\ }\textbf {\bibinfo {volume} {17}},\ \bibinfo {pages} {2300075} (\bibinfo {year} {2023})}\BibitemShut {NoStop}%
\bibitem [{\citenamefont {Conforti}\ and\ \citenamefont {Trillo}(2013)}]{conforti_dispersive_2013}%
  \BibitemOpen
  \bibfield  {author} {\bibinfo {author} {\bibfnamefont {M.}~\bibnamefont {Conforti}}\ and\ \bibinfo {author} {\bibfnamefont {S.}~\bibnamefont {Trillo}},\ }\bibfield  {title} {\bibinfo {title} {Dispersive wave emission from wave breaking},\ }\href@noop {} {\bibfield  {journal} {\bibinfo  {journal} {Optics Letters}\ }\textbf {\bibinfo {volume} {38}},\ \bibinfo {pages} {3815} (\bibinfo {year} {2013})}\BibitemShut {NoStop}%
\bibitem [{\citenamefont {Jang}\ \emph {et~al.}(2014)\citenamefont {Jang}, \citenamefont {Erkintalo}, \citenamefont {Murdoch},\ and\ \citenamefont {Coen}}]{jang_observation_2014}%
  \BibitemOpen
  \bibfield  {author} {\bibinfo {author} {\bibfnamefont {J.~K.}\ \bibnamefont {Jang}}, \bibinfo {author} {\bibfnamefont {M.}~\bibnamefont {Erkintalo}}, \bibinfo {author} {\bibfnamefont {S.~G.}\ \bibnamefont {Murdoch}},\ and\ \bibinfo {author} {\bibfnamefont {S.}~\bibnamefont {Coen}},\ }\bibfield  {title} {\bibinfo {title} {Observation of dispersive wave emission by temporal cavity solitons},\ }\href@noop {} {\bibfield  {journal} {\bibinfo  {journal} {Optics Letters}\ }\textbf {\bibinfo {volume} {39}},\ \bibinfo {pages} {5503} (\bibinfo {year} {2014})}\BibitemShut {NoStop}%
\bibitem [{\citenamefont {Goda}\ and\ \citenamefont {Jalali}(2013)}]{goda_dispersive_2013}%
  \BibitemOpen
  \bibfield  {author} {\bibinfo {author} {\bibfnamefont {K.}~\bibnamefont {Goda}}\ and\ \bibinfo {author} {\bibfnamefont {B.}~\bibnamefont {Jalali}},\ }\bibfield  {title} {\bibinfo {title} {Dispersive {Fourier} transformation for fast continuous single-shot measurements},\ }\href {https://doi.org/10.1038/nphoton.2012.359} {\bibfield  {journal} {\bibinfo  {journal} {Nature Photonics}\ }\textbf {\bibinfo {volume} {7}},\ \bibinfo {pages} {102} (\bibinfo {year} {2013})}\BibitemShut {NoStop}%
\bibitem [{\citenamefont {Godin}\ \emph {et~al.}(2022)\citenamefont {Godin}, \citenamefont {Sader}, \citenamefont {Khodadad~Kashi}, \citenamefont {Hanzard}, \citenamefont {Hideur}, \citenamefont {Moss}, \citenamefont {Morandotti}, \citenamefont {Genty}, \citenamefont {Dudley}, \citenamefont {Pasquazi}, \citenamefont {Kues},\ and\ \citenamefont {Wetzel}}]{godin_recent_2022}%
  \BibitemOpen
  \bibfield  {author} {\bibinfo {author} {\bibfnamefont {T.}~\bibnamefont {Godin}}, \bibinfo {author} {\bibfnamefont {L.}~\bibnamefont {Sader}}, \bibinfo {author} {\bibfnamefont {A.}~\bibnamefont {Khodadad~Kashi}}, \bibinfo {author} {\bibfnamefont {P.-H.}\ \bibnamefont {Hanzard}}, \bibinfo {author} {\bibfnamefont {A.}~\bibnamefont {Hideur}}, \bibinfo {author} {\bibfnamefont {D.~J.}\ \bibnamefont {Moss}}, \bibinfo {author} {\bibfnamefont {R.}~\bibnamefont {Morandotti}}, \bibinfo {author} {\bibfnamefont {G.}~\bibnamefont {Genty}}, \bibinfo {author} {\bibfnamefont {J.~M.}\ \bibnamefont {Dudley}}, \bibinfo {author} {\bibfnamefont {A.}~\bibnamefont {Pasquazi}}, \bibinfo {author} {\bibfnamefont {M.}~\bibnamefont {Kues}},\ and\ \bibinfo {author} {\bibfnamefont {B.}~\bibnamefont {Wetzel}},\ }\bibfield  {title} {\bibinfo {title} {Recent advances on time-stretch dispersive {Fourier} transform and its applications},\ }\href {https://doi.org/10.1080/23746149.2022.2067487} {\bibfield  {journal} {\bibinfo  {journal}
  {Advances in Physics: X}\ }\textbf {\bibinfo {volume} {7}},\ \bibinfo {pages} {2067487} (\bibinfo {year} {2022})}\BibitemShut {NoStop}%
\bibitem [{\citenamefont {Kolner}(1994)}]{kolner_space-time_1994}%
  \BibitemOpen
  \bibfield  {author} {\bibinfo {author} {\bibfnamefont {B.}~\bibnamefont {Kolner}},\ }\bibfield  {title} {\bibinfo {title} {Space-time duality and the theory of temporal imaging},\ }\href {https://doi.org/10.1109/3.301659} {\bibfield  {journal} {\bibinfo  {journal} {IEEE Journal of Quantum Electronics}\ }\textbf {\bibinfo {volume} {30}},\ \bibinfo {pages} {1951} (\bibinfo {year} {1994})}\BibitemShut {NoStop}%
\bibitem [{\citenamefont {Chou}\ \emph {et~al.}(2008)\citenamefont {Chou}, \citenamefont {Solli},\ and\ \citenamefont {Jalali}}]{chou_real-time_2008}%
  \BibitemOpen
  \bibfield  {author} {\bibinfo {author} {\bibfnamefont {J.}~\bibnamefont {Chou}}, \bibinfo {author} {\bibfnamefont {D.~R.}\ \bibnamefont {Solli}},\ and\ \bibinfo {author} {\bibfnamefont {B.}~\bibnamefont {Jalali}},\ }\bibfield  {title} {\bibinfo {title} {Real-time spectroscopy with subgigahertz resolution using amplified dispersive {Fourier} transformation},\ }\bibfield  {journal} {\bibinfo  {journal} {Applied Physics Letters}\ }\textbf {\bibinfo {volume} {92}},\ \href {https://doi.org/10.1063/1.2896652} {10.1063/1.2896652} (\bibinfo {year} {2008}),\ \bibinfo {note} {publisher: AIP Publishing}\BibitemShut {NoStop}%
\bibitem [{\citenamefont {Solli}\ \emph {et~al.}(2008)\citenamefont {Solli}, \citenamefont {Ropers},\ and\ \citenamefont {Jalali}}]{solli_active_2008}%
  \BibitemOpen
  \bibfield  {author} {\bibinfo {author} {\bibfnamefont {D.~R.}\ \bibnamefont {Solli}}, \bibinfo {author} {\bibfnamefont {C.}~\bibnamefont {Ropers}},\ and\ \bibinfo {author} {\bibfnamefont {B.}~\bibnamefont {Jalali}},\ }\bibfield  {title} {\bibinfo {title} {Active {Control} of {Rogue} {Waves} for {Stimulated} {Supercontinuum} {Generation}},\ }\bibfield  {journal} {\bibinfo  {journal} {Physical Review Letters}\ }\textbf {\bibinfo {volume} {101}},\ \href {https://doi.org/10.1103/physrevlett.101.233902} {10.1103/physrevlett.101.233902} (\bibinfo {year} {2008}),\ \bibinfo {note} {publisher: American Physical Society (APS)}\BibitemShut {NoStop}%
\bibitem [{\citenamefont {Wetzel}\ \emph {et~al.}(2012)\citenamefont {Wetzel}, \citenamefont {Stefani}, \citenamefont {Larger}, \citenamefont {Lacourt}, \citenamefont {Merolla}, \citenamefont {Sylvestre}, \citenamefont {Kudlinski}, \citenamefont {Mussot}, \citenamefont {Genty}, \citenamefont {Dias},\ and\ \citenamefont {Dudley}}]{wetzel_real-time_2012}%
  \BibitemOpen
  \bibfield  {author} {\bibinfo {author} {\bibfnamefont {B.}~\bibnamefont {Wetzel}}, \bibinfo {author} {\bibfnamefont {A.}~\bibnamefont {Stefani}}, \bibinfo {author} {\bibfnamefont {L.}~\bibnamefont {Larger}}, \bibinfo {author} {\bibfnamefont {P.~A.}\ \bibnamefont {Lacourt}}, \bibinfo {author} {\bibfnamefont {J.~M.}\ \bibnamefont {Merolla}}, \bibinfo {author} {\bibfnamefont {T.}~\bibnamefont {Sylvestre}}, \bibinfo {author} {\bibfnamefont {A.}~\bibnamefont {Kudlinski}}, \bibinfo {author} {\bibfnamefont {A.}~\bibnamefont {Mussot}}, \bibinfo {author} {\bibfnamefont {G.}~\bibnamefont {Genty}}, \bibinfo {author} {\bibfnamefont {F.}~\bibnamefont {Dias}},\ and\ \bibinfo {author} {\bibfnamefont {J.~M.}\ \bibnamefont {Dudley}},\ }\bibfield  {title} {\bibinfo {title} {Real-time full bandwidth measurement of spectral noise in supercontinuum generation},\ }\bibfield  {journal} {\bibinfo  {journal} {Scientific Reports}\ }\textbf {\bibinfo {volume} {2}},\ \href {https://doi.org/10.1038/srep00882} {10.1038/srep00882}
  (\bibinfo {year} {2012}),\ \bibinfo {note} {publisher: Springer Science and Business Media LLC}\BibitemShut {NoStop}%
\bibitem [{\citenamefont {Cutrona}\ \emph {et~al.}(2023)\citenamefont {Cutrona}, \citenamefont {Cecconi}, \citenamefont {Hanzard}, \citenamefont {Rowley}, \citenamefont {Das}, \citenamefont {Cooper}, \citenamefont {Peters}, \citenamefont {Olivieri}, \citenamefont {Wetzel}, \citenamefont {Morandotti}, \citenamefont {Chu}, \citenamefont {Little}, \citenamefont {Moss}, \citenamefont {Totero~Gongora}, \citenamefont {Peccianti},\ and\ \citenamefont {Pasquazi}}]{cutrona_nonlocal_2023}%
  \BibitemOpen
  \bibfield  {author} {\bibinfo {author} {\bibfnamefont {A.}~\bibnamefont {Cutrona}}, \bibinfo {author} {\bibfnamefont {V.}~\bibnamefont {Cecconi}}, \bibinfo {author} {\bibfnamefont {P.~H.}\ \bibnamefont {Hanzard}}, \bibinfo {author} {\bibfnamefont {M.}~\bibnamefont {Rowley}}, \bibinfo {author} {\bibfnamefont {D.}~\bibnamefont {Das}}, \bibinfo {author} {\bibfnamefont {A.}~\bibnamefont {Cooper}}, \bibinfo {author} {\bibfnamefont {L.}~\bibnamefont {Peters}}, \bibinfo {author} {\bibfnamefont {L.}~\bibnamefont {Olivieri}}, \bibinfo {author} {\bibfnamefont {B.}~\bibnamefont {Wetzel}}, \bibinfo {author} {\bibfnamefont {R.}~\bibnamefont {Morandotti}}, \bibinfo {author} {\bibfnamefont {S.~T.}\ \bibnamefont {Chu}}, \bibinfo {author} {\bibfnamefont {B.~E.}\ \bibnamefont {Little}}, \bibinfo {author} {\bibfnamefont {D.~J.}\ \bibnamefont {Moss}}, \bibinfo {author} {\bibfnamefont {J.~S.}\ \bibnamefont {Totero~Gongora}}, \bibinfo {author} {\bibfnamefont {M.}~\bibnamefont {Peccianti}},\ and\ \bibinfo {author} {\bibfnamefont
  {A.}~\bibnamefont {Pasquazi}},\ }\bibfield  {title} {\bibinfo {title} {Nonlocal bonding of a soliton and a blue-detuned state in a microcomb laser},\ }\bibfield  {journal} {\bibinfo  {journal} {Communications Physics}\ }\textbf {\bibinfo {volume} {6}},\ \href {https://doi.org/10.1038/s42005-023-01372-0} {10.1038/s42005-023-01372-0} (\bibinfo {year} {2023}),\ \bibinfo {note} {publisher: Springer Science and Business Media LLC}\BibitemShut {NoStop}%
\bibitem [{\citenamefont {Lapre}\ \emph {et~al.}(2020)\citenamefont {Lapre}, \citenamefont {Billet}, \citenamefont {Meng}, \citenamefont {Genty},\ and\ \citenamefont {Dudley}}]{lapre_dispersive_2020}%
  \BibitemOpen
  \bibfield  {author} {\bibinfo {author} {\bibfnamefont {C.}~\bibnamefont {Lapre}}, \bibinfo {author} {\bibfnamefont {C.}~\bibnamefont {Billet}}, \bibinfo {author} {\bibfnamefont {F.}~\bibnamefont {Meng}}, \bibinfo {author} {\bibfnamefont {G.}~\bibnamefont {Genty}},\ and\ \bibinfo {author} {\bibfnamefont {J.~M.}\ \bibnamefont {Dudley}},\ }\bibfield  {title} {\bibinfo {title} {Dispersive {Fourier} transform characterization of multipulse dissipative soliton complexes in a mode-locked soliton-similariton laser},\ }\href {https://doi.org/10.1364/osac.384501} {\bibfield  {journal} {\bibinfo  {journal} {OSA Continuum}\ }\textbf {\bibinfo {volume} {3}},\ \bibinfo {pages} {275} (\bibinfo {year} {2020})},\ \bibinfo {note} {publisher: Optica Publishing Group}\BibitemShut {NoStop}%
\bibitem [{\citenamefont {Copie}\ \emph {et~al.}(2017{\natexlab{b}})\citenamefont {Copie}, \citenamefont {Conforti}, \citenamefont {Kudlinski}, \citenamefont {Trillo},\ and\ \citenamefont {Mussot}}]{Copie_mod_2017}%
  \BibitemOpen
  \bibfield  {author} {\bibinfo {author} {\bibfnamefont {F.}~\bibnamefont {Copie}}, \bibinfo {author} {\bibfnamefont {M.}~\bibnamefont {Conforti}}, \bibinfo {author} {\bibfnamefont {A.}~\bibnamefont {Kudlinski}}, \bibinfo {author} {\bibfnamefont {S.}~\bibnamefont {Trillo}},\ and\ \bibinfo {author} {\bibfnamefont {A.}~\bibnamefont {Mussot}},\ }\bibfield  {title} {\bibinfo {title} {Modulation instability in the weak dispersion regime of a dispersion modulated passive fiber-ring cavity},\ }\href {https://doi.org/10.1364/OE.25.011283} {\bibfield  {journal} {\bibinfo  {journal} {Opt. Express}\ }\textbf {\bibinfo {volume} {25}},\ \bibinfo {pages} {11283} (\bibinfo {year} {2017}{\natexlab{b}})}\BibitemShut {NoStop}%
\end{thebibliography}%

\end{document}